\documentclass[galaxies,review,accept,pdftex,oneauthor,a4paper]{Definitions/mdpi} 

\firstpage{1} 
\pubvolume{11}
\issuenum{1}
\articlenumber{59}
\pubyear{2023}
\copyrightyear{2023}
\externaleditor{Academic Editors: Vladimir Korchagin, Maxim Yu.\ Khlopov, Orchidea Maria Lecian}
\datereceived{4 March 2023}
\daterevised{13 April 2023} 
\dateaccepted{14 April 2023}
\datepublished{18 April 2023} \hreflink{https://doi.org/10.3390/galaxies11020059} 

\Title{The effect of the LMC on the Milky Way system}

\TitleCitation{LMC and the Milky Way}

\Author{Eugene Vasiliev \orcidA{}}

\AuthorNames{Eugene Vasiliev}

\AuthorCitation{Vasiliev, E.}

\address[1]{\quad Institute of Astronomy, Madingley road, Cambridge, UK, CB30HA; eugvas@protonmail.com}

\abstract{
We review the recent theoretical and observational developments concerning the interaction of the Large Magellanic Cloud (LMC) with the Milky Way and its neighbourhood. An emerging picture is that the LMC is a fairly massive companion (10--20\% of the Milky Way mass) and just passed the pericentre of its orbit, likely for the first time. 
The gravitational perturbation caused by the LMC is manifested at different levels. 
The most immediate effect is the deflection of orbits of stars, stellar streams or satellite galaxies passing in the vicinity of the LMC. Less well known but equally important is the displacement (reflex motion) of central regions of the Milky Way about the centre of mass of both galaxies. Since the Milky Way is not a rigid body, this displacement varies with the distance from the LMC, and as a result, the Galaxy is deformed and its outer regions (beyond a few tens kpc) acquire a net velocity with respect to its centre. These phenomena need to be taken into account at the level of precision warranted by current and future observational data, and improvements on the modelling side are also necessary for an adequate interpretation of these data.
}


\usepackage{xspace}
\newcommand{\kms}{km\:s$^{-1}$\xspace}

\newcommand{\new}[1]{#1}

\begin{document}

\section{Introduction}

The Magellanic Clouds are the best known and the largest galaxies in the Milky Way neighbourhood, and the only ones prominently visible on the night sky. They were always believed to be Galactic satellites, but the views on how much they affect the Milky Way has changed over years. This review summarises the recent observational and theoretical developments that together point to a large impact of the LMC on the dynamical processes in the outer Galaxy and even beyond. 

The long-accepted view that the Clouds have been circling around our Galaxy for many orbits (e.g., \citep{Tremaine1976, Murai1980, Lin1982,Gardiner1994}) gradually gave way to the now-standard picture that their past orbits are highly eccentric with apocentres well beyond 200~kpc and orbital periods in excess of 5~Gyr, or even unbound \citep{Besla2007}. This conclusion is largely driven by the high measured tangential velocity component of the LMC \citep{Kallivayalil2006a,Kallivayalil2013}, but would also naturally explain the continuous presence of the Small Magellanic Cloud (SMC) and other more recently discovered satellites of the LMC, which would have been stripped by the Milky Way tidal field if the Clouds were on a much tighter orbit around the Galaxy. Nevertheless, as discussed in Section~\ref{sec:orbit}, there remain large uncertainties in the past orbit of the LMC and a number of caveats in reconstructing it.

Another recent revelation was the mounting evidence (reviewed in Section~\ref{sec:lmc_mass}) that the LMC is fairly massive, only 5--10 times smaller than the Milky Way itself. Interestingly, more than 50 years ago a similarly titled paper by \citet{Avner1967} adopted a 1:10 mass ratio, but back in the days both galaxies were believed to be 10 times less massive than current estimates. 
The high mass of the LMC naturally explains a number of independent dynamical effects discovered recently, which are the major part of this review and are discussed in Section~\ref{sec:dynamics}. 

\section{Introducing the participants}

Before proceeding to the discussion of various aspects of the Milky Way--LMC interaction, we summarise the current knowledge about our Galaxy and the Magellanic system, in as much as it is relevant for their interaction.

\subsection{Milky Way}  \label{sec:milkyway}

Naturally, the Galaxy that we live in happens to be the best studied one. The quantity and precision of observational data have skyrocketed in the last decade with the advent of large-scale spectroscopic surveys such as SEGUE \citep{Yanny2009}, APOGEE \citep{Majewski2017} and LAMOST \citep{Zhao2012}, and since 2018, the \textit{Gaia} satellite provides positions, parallaxes and proper motion (PM) measurements for $\gtrsim1.5\times10^9$ stars across the entire sky \citep{Gaia2018, Gaia2021, Gaia2023}. 

One of the major consequences of this wealth of data is that now it becomes possible to explore not only the global structure of the Galaxy, but also relatively small ($\sim 10-20\%$) deviations from equilibrium. The best-known example is the so-called ``Gaia snail'' -- a winding spiral pattern in the $\{z,v_z\}$ phase space \citep{Antoja2018}, indicating ongoing phase mixing in a perturbed disc. The leading (though not universally accepted) theory is that it was created by a passage of a massive satellite (likely the Sagittarius dSph) through the disc plane $\sim 1$~Gyr ago \citep[e.g.,][]{Widrow2012,Laporte2018b}. As will be discussed in this review, the LMC is unlikely to significantly disturb the inner Galaxy, but creates a comparable if not larger perturbation in the outer halo.

Our knowledge of Galactic structure and mass distribution is mostly derived from measuring the kinematics of stars and other tracers, coupled with suitable modelling procedures. The observational data on the kinematics of tracers are abundant in the Solar neighbourhood, but increasingly scarce and less precise at large distances. For instance, the line-of-sight velocity of individual stars is usually measured to within a few \kms, but the uncertainty in the transverse velocity at 50~kpc is at the level of 100~\kms in the current \textit{Gaia} DR3, comparable to the velocity dispersion in the halo. Much better precision can be obtained for globular clusters and satellite galaxies by averaging the PM of many stars, but these tracers themselves are not very numerous, with only a few dozen galaxies and a handful of clusters beyond 50~kpc. 
Last but not least, stellar streams formed by tidal stripping of stars from clusters and satellites (including objects that by now are completely disrupted) are very valuable probes of Galactic potential, since stars in each stream approximately follow the orbit of its progenitor and span a large range of Galactocentric distances. Refs.\ \citep{Eyre2011, Sanders2013, Bovy2014, Gibbons2014} and other studies discuss the mechanics of stream formation and their use for constraining the Galactic potential. Nearly a hundred streams in the Milky Way have been catalogued so far \citep{Mateu2023}, although only a few reach large enough Galactocentric distances to be affected by the LMC. Of these, the most important are the Sagittarius stream \citep{Majewski2003} produced by the eponimous galaxy (one of the closest and most massive Milky Way satellites), and the Orphan--Chenab stream without a known progenitor \citep{Grillmair2006,Belokurov2007,Shipp2018,Koposov2019}. Both span more than $180^\circ$ on the sky and tens of kpc in distance, and are sensitive to the LMC perturbation, as discussed below.

Many of the ``classical'' modelling techniques such as Jeans equations or distribution function-based models rely on the assumption of dynamical equilibrium, and need a sufficient number of individual tracers to exploit this assumption. Any single kinematic tracer provides very weak constraints (essentially, only a lower limit on the escape velocity, assuming that the object is bound to the Galaxy), and a joint distribution of many objects is needed to exploit the assumption of phase-mixedness built into these modelling approaches. On the other hand, even a single stream can provide interesting constraints on the potential in the range of radii spanned by its track \citep[e.g.,][]{Johnston1999,Bonaca2018}.
\citet{BlandHawthorn2016} give a comprehensive review of Milky Way structure at all scales in the pre-\textit{Gaia} era, and \citet{Wang2020} summarize the more recent developments in determining the total mass of our Galaxy and its spatial distribution. The precision and scatter in these measurements greatly vary with Galactocentric distance. Naturally, the best-studied region is around the Sun ($\sim 8$~kpc), where the enclosed mass is known to within $\lesssim 10\%$ and is of order $10^{11}\,M_\odot$. On the other hand, at larger distances the constraints are generally weaker and display a large scatter between different studies. Sometimes the well-measured mass distribution in the inner Galaxy is extrapolated to large distances assuming a particular functional form (e.g., a Navarro--Frenk--White profile), resulting in unrealistically small uncertaities in absense of actual kinematic tracers. An emerging consensus is that the \new{stellar and} total mass of the Milky Way are of order \new{$5\times10^{10}\,M_\odot$ and} $10^{12}\,M_\odot$ \new{respectively}, with an optimistic level of uncertainty being $\sim20-30\%$. The mass enclosed within 50~kpc is better constrained 
and is at the level $4\times 10^{11}\,M_\odot$, corresponding to a circular velocity of 185~\kms. \new{On the other hand, the 3d shape of the Galactic halo remains largely unknown, with different studies arriving at conflicting results. }

\subsection{LMC}  \label{sec:lmc}

\subsubsection{Mass}  \label{sec:lmc_mass}

\begin{figure}
\input{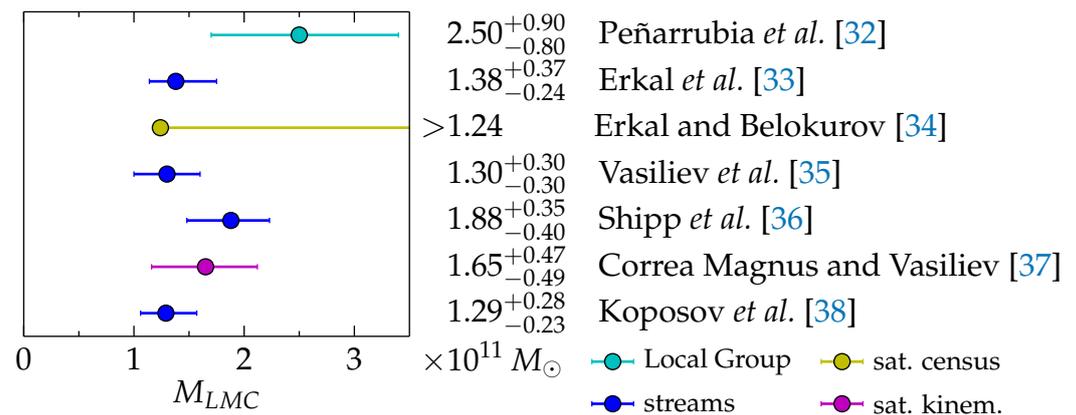}
\caption{LMC mass estimates from different methods: blue -- perturbations inflicted on stellar streams \citep{Erkal2019, Vasiliev2021, Shipp2021, Koposov2023}; cyan -- momentum balance in the Local Group \citep{Penarrubia2016}; yellow -- census of LMC satellites \citep{Erkal2020a}; magenta -- kinematics of Milky Way satellites \citep{CorreaMagnus2022}.}  
\label{fig:lmc_mass}
\end{figure}

The LMC is undoubtedly the largest satellite of the Milky Way, but just how large it is only became apparent relatively recently (see Figure~\ref{fig:lmc_mass} for a compilation of estimates). 
The stellar mass of the LMC is $\sim 3\times10^9\,M_\odot$ \citep{vanderMarel2002} and gas mass $\sim 5\times10^8\,M_\odot$ \citep{Fox2014}, but its total mass ought to be significantly larger. There are several lines of evidence suggesting that the LMC mass is in the range $(1-2)\times 10^{11}\,M_\odot$.
\begin{itemize}
\item Cosmological stellar mass--halo mass relation.
\item Mass modelling from internal kinematics.
\item Census of its satellites.
\item Interaction with the SMC.
\item Dynamical perturbation of stellar streams.
\item Kinematic and spatial distortions in the Milky Way halo.
\item Hubble flow in the Local Universe.
\end{itemize}

Let us consider these arguments in turn. 
In the standard cosmological scenario, the baryonic mass of the LMC is expected to comprise about 1\% of its total mass \citep{Moster2013, Read2019}, implying the latter to be $\gtrsim 10^{11}\,M_\odot$. The caveat is that if the LMC has been orbiting the Milky Way for some time, its dark halo would have been tidally stripped, similarly to what has happened to the Sagittarius dSph, whose present-day total mass is only a few times higher than its stellar mass \cite{Vasiliev2020}. However, as discussed below, this scenario appears to be unlikely.

The three-dimensional structure and kinematics of stars in the LMC have been extensively studied in several large-scale photometric, spectroscopic and astrometric surveys, and in combination with suitable modelling techniques making it possible to determine its total mass distribution. Using the \textit{Hubble} space telescope (\textit{HST}) astrometry, \citep{vanderMarel2014} measured the asymptotic circular velocity to be $\sim 90$~\kms at the distance of 9~kpc, implying an enclosed mass of $\sim 2\times10^{10}\,M_\odot$. More recent analysis based on \textit{Gaia} DR2 \citep{Vasiliev2018,Wan2020} and (e)DR3 \citep{Luri2021,JimenezArranz2023} corroborates these findings. The mass profile at larger distances is difficult to measure, not only due to the dearth of kinematic tracers, but also due to apparent disturbances in the stellar motions likely caused by the interaction with the SMC \cite{Choi2022,Cullinane2022}. Nevertheless, the measured peak circular velocity translates to the halo mass of order $10^{11}\,M_\odot$ \citep{Katz2019}, in agreement with the stellar mass-based estimate (a consequence of the baryonic Tully--Fisher relation).

The LMC possesses a number of satellites, as discussed in the next section, but in order to hold them up, it needs to be massive enough. In particular, the relative velocity of the LMC--SMC pair ($\sim 100$--120~\kms) is quite high, and would correspond to an unbound orbit if their total mass is $\lesssim 10^{11}\,M_\odot$ \citep{Kallivayalil2013}. It is considered unlikely that the two Clouds just happened to be both coming into the vicinity of the Milky Way independently, without being a bound system.

Perturbations of stellar streams and other kinematic tracers in the outer Milky Way halo are discussed at length in Section~\ref{sec:dynamics}, but they all are best explained if the LMC mass is in the range (1--2)${}\times10^{11}\,M_\odot$. 
Finally, \citet{Penarrubia2016} estimated that an LMC mass of up to $2.5\times10^{11}\,M_\odot$ is needed to best match the local Hubble flow (at a distances of a few Mpc); this argument is discussed at the end of Section~\ref{sec:local_group}.

\subsubsection{Satellites}  \label{sec:lmc_satellites}

\begin{table}
\caption{Possible satellites of the LMC, according to several studies:
\citet{Jethwa2016}, \citet{Sales2017}, \citet{Kallivayalil2018}, \citet{Erkal2020a}, \citet{Patel2020}, \citet{Battaglia2022}, \citet{CorreaMagnus2022}. Highly likely, uncertain, unlikely and recently captured/interacted satellites are denoted by +, ?, -- and $c$ respectively. Objects in brackets lacked line-of-sight velocity measurements at the time of writing of the corresponding paper, but are consistent with being satellites for some values of velocity (note that all objects in J16 and S17 also lacked PM measurements).
} \label{tab:LMCsatellites}
\begin{tabular}{lccccccc}
\hline
galaxy        & J16 & S17 & K18 & E20 & P20 & B22 & C22 \\
\hline
Carina        &     &     &     &  -- &  -- &  ?  &  -- \\
Carina~II     &     &     &  +  &  +  &  +  &  +  &  +  \\
Carina~III    &     &     &  +  &  +  &  +  &  +  &  +  \\
Delve~2       &     &     &     &     &     &     & (?) \\
Eridanus~III  &     & (?) &     & (?) &     &     & (?) \\
Grus~I        &  ?  &  ?  &  -- &  -- &     &  -- &  -- \\
Grus~II       & (?) &     &     & (--)&     & $c$ & $c$ \\
Horologium~I  &  +  &  +  &  +  &  +  &  +  &  +  &  +  \\
Horologium~II & (+) & (?) &     & (--)&     &  ?  &  ?  \\
Hydrus~I      &     &     &  +  &  +  &  +  &  +  &  +  \\
Pictor~I      & (?) &     &     & (--)&     &     & (--)\\
Pictor~II     &     &     &     &     &     &     & (?) \\
Phoenix~II    & (+) & (?) & (?) &  +  & $c$ &  +  &  +  \\
Reticulum~II  &  +  &  ?  &  -- &  +  & $c$ &  +  &  ?  \\
Reticulum~III & (?) & (?) &     & (--)&     &  -- &  -- \\
SMC           &     &     &     &  +  &     &     &  +  \\
Tucana~II     &  +  &  ?  &  -- &  -- &     &  -- &  -- \\
Tucana~III    & (?) &     &  -- &  -- &  -- &  -- &  -- \\
Tucana~IV     & (+) & (?) &     & (--)&     &  ?  & $c$ \\
Tucana~V      & (+) & (?) &     &     &     &  -- &  -- \\
\hline
\end{tabular}
\end{table}

The LMC comes not alone, but with an entourage of its own satellite galaxies. One of them, the SMC, is well known, whereas other, much fainter ones, have been discovered relatively recently, mostly in the Dark Energy Survey \citep{Koposov2015,Bechtol2015,DrlicaWagner2015}. Several studies have estimated the probability of being a satellite of the LMC for a number of classical and ultrafaint dwarf galaxies; Table~\ref{tab:LMCsatellites} summarizes these findings. SMC and 5--6 ultrafaint dwarfs have been universally accepted as members of the LMC cohort, while a couple of other galaxies have experienced a recent interaction with the LMC, but are unlikely to be its satellites in the past. By comparing this number of satellites with cosmological simulations, \citet{Erkal2020a} estimated the LMC mass to be at least $1.2\times10^{11}\,M_\odot$, but noted that many more even fainter satellites might await detection (see also \citep{Jethwa2016}). Naturally, the association of these galaxies with the LMC is conditioned on the LMC mass (see Figure~13 in \citep{CorreaMagnus2022}), but for most of these candidate satellites it stays fairly high in the entire range of reasonable masses.

\subsection{SMC}  \label{sec:SMC}

The SMC, being in conspicuous proximity to the LMC and only a few times less luminous, was long suspected to be a satellite of the latter. Reconstruction of its 3d shape from photometry of red clump stars and other standard candles indicates that it is fairly extended along the line of sight: with a mean distance of $\sim 68$~kpc, its thickness is of order 10~kpc (see Table~1 in \citep{Tatton2021} for a summary of studies). This, together with kinematic evidence for a strong perturbation, suggests a rather dramatic harassment by its bigger sibling. The most recent close encounter between the two Clouds occurred around 150~Myr ago at a distance of $\lesssim 10$~kpc \citep{Besla2012,Kallivayalil2013,Zivick2018}, but in all likelihood, they have orbited each other for several Gyr in the past. As mentioned earlier, the requirement that the Clouds form a bound pair imposes a lower limit on their total mass of order $10^{11}\,M_\odot$. It is quite likely that the SMC had been much more massive in the past than it is now, having been stripped of most of its dark matter halo and a considerable fraction of its gas. Its present-day mass is estimated to be $\mathcal O(10^{10})\,M_\odot$, thus it has negligible effect on the Milky Way, but still an appreciable effect on the LMC. In particular, if both galaxies should be considered as a binary system, then the reconstruction of their past orbits should be performed for the centre-of-mass velocity rather than the LMC velocity.

\subsection{Magellanic stream, bridge and other structures}  \label{sec:bridge}

A long and prominent gas stream trailing behind the LMC was discovered by \citet{Mathewson1974}. In the earlier picture where the LMC was orbiting the Milky Way for a long time, the stream was believed to be created either by the ram pressure stripping or the tidal force of our Galaxy \citep{Murai1980}. However, in the more recent scenario of the first approach, its origin must be different, and the prevailing view is that it is formed primarily from the gas stripped from the SMC while it orbits the LMC \citep{Besla2010}. \new{The leading arm of this gas stream is more difficult to explain in this scenario, especially when including the hot gas corona of the LMC \citep{TepperGarcia2019}, though the debate about the role of hot corona is still open \citep{Lucchini2020}.} \citet{DOnghia2016} present a comprehensive review of the Magellanic gas stream.

Unlike the Sagittarius galaxy, the LMC does not produce any noticeable stellar stream, further strengthening the argument that it is likely on its first passage and/or embedded in a massive protective dark halo envelope. However, recently a tentative evidence for a stellar counterpart of the gas stream was provided by \citet{Zaritsky2020} and \citet{Petersen2022} in the trailing and leading arms, respectively. \new{The measured microlensing optical depth towards the LMC can be explained by the lensing of the faint stellar counterpart of the Magellanic stream by stars of the LMC halo \citep{Besla2013}.} A young star cluster was discovered  by \citet{PriceWhelan2019} near the leading arm, but follow-up spectroscopic observations disagreed about its possible association with the stream: \citet{Nidever2019} found it likely while \citet{Bellazzini2019} placed it on a much tighter orbit around the Galaxy with a pericentre $\lesssim 15$~kpc and a period of $\sim 0.5$~Gyr.

On the other hand, there are numerous stellar structures in the region between the two Clouds (the Magellanic bridge) \citep{Belokurov2017}, likely formed out of stars stripped from both galaxies in the recent collision \citep{Belokurov2019a,Zivick2019}, although some perturbations in the outer disc of the LMC are more likely to be caused by the Milky Way tidal force \citep{Mackey2016,Cullinane2022}.

\section{Orbit of the LMC}  \label{sec:orbit}

\subsection{Present-day position and velocity}  \label{sec:present_day_posvel}

\begin{table}
\caption{Compilation of measurements of the LMC PM. $\alpha_0$ and $\delta_0$ are the ICRS coordinates of the measured centre point (or adopted from another study, shown in brackets); $\mu_{\alpha*}\equiv ({\rm d}\alpha/{\rm d}t)\cos\delta$, $\mu_\delta\equiv {\rm d}\delta/{\rm d}t$ are the corresponding PM components in mas\,yr$^{-1}$.  See also Table~2 in \citet{DOnghia2016} for a summary of older measurements.
}  \label{tab:LMCposvel}
\begin{tabular}{lccll}
\hline
reference & $\alpha_0$ & $\delta_0$ & $\mu_{\alpha*}$ & $\mu_\delta$ \\
\hline
\citet{Kallivayalil2006a}&($81.90^\circ$)&($-69.87^\circ$)& $2.03\phantom{0}\pm 0.08$ & $0.44\phantom{0}\pm 0.05$ \\
\citet{vanderMarel2014}  & $78.76^\circ$ & $-69.19^\circ$ & $1.910 \pm 0.02$ & $0.229 \pm 0.05$ \\
\citet{Helmi2018}        &($78.77^\circ$)&($-69.01^\circ$)& $1.850 \pm 0.03$ & $0.234 \pm 0.03$ \\
--- (alt.centre)         &($81.28^\circ$)&($-69.78^\circ$)& $1.890 \pm 0.03$ & $0.314 \pm 0.03$ \\
\citet{Wan2020}          & $80.90^\circ$ & $-68.74^\circ$ & $1.878 \pm 0.03$ & $0.293 \pm 0.03$ \\
\citet{Luri2021}         &($81.28^\circ$)&($-69.78^\circ$)& $1.858 \pm 0.02$ & $0.385 \pm 0.02$ \\
--- (alt.centre)         & $81.07^\circ$ & $-69.41^\circ$ & $1.847 \pm 0.02$ & $0.371 \pm 0.02$ \\
\hline
\end{tabular}
\end{table}

The LMC is currently located at a distance of $\sim 49.6\pm0.5$~kpc \citep{Pietrzynski2019}, its heliocentric line-of-sight velocity is $\sim 262.2\pm3.4$~\kms \citep{vanderMarel2002}, and its PM has been measured with an increasing precision first by \textit{HST} \citep{Kallivayalil2006a,Kallivayalil2013} and more recently by \textit{Gaia} \citep{Helmi2018,Luri2021}. 
There is some ambiguity regarding the position of the LMC centre (see section 4.1 in \citet{vanderMarel2014} for a discussion), and different centre locations correspond to different values of mean PM \citep{Helmi2018,Wan2020,Luri2021} due to both perspective effects and internal motion in the LMC. The line-of-sight velocity also has a significant spatial gradient of $\sim10$~\kms per degree (mostly with declination $\delta$), but unfortunately this factor is ignored in almost all studies, which adopt the same value but attribute it to a different centre point. 
Table~\ref{tab:LMCposvel} lists a number of recent measurements of the mean PM and the associated centre locations (in some studies it was kept fixed while in others it was varied during the fit). 
The LMC moves away from the Milky Way with a Galactocentric radial velocity of $\sim 70$~\kms, whereas its tangential velocity is around 310~\kms (well above the circular velocity at that distance). These values imply that it just recently passed the pericentre of its orbit, but the reconstruction of its orbit further into the past depends on a number of factors, most importantly the masses of both galaxies.

Until the more precise space-based measurements of the LMC velocity became available, it was commonly assumed that it had completed several orbits around the Milky Way over the Hubble time \citep[e.g.,][]{Murai1980}. However, the high transverse velocity measured by \citet{Kallivayalil2006a} implies a very eccentric or even hyperbolic orbit, unless the Milky Way is very massive. (Note however that the more recent PM measurements reduce the tangential velocity by a few tens \kms, weakening the evidence for a highly eccentric or unbound orbit). Thus \citet{Besla2007} put forward the now favourite scenario of a first infall (i.e., no previous pericentre passages within the last 10~Gyr, \new{although the precise definition varies between studies, e.g., \citet{Patel2020} denoted first-infall orbits as having a period above 6~Gyr and apocentre exceeding the virial radius of the Milky Way}). 
There are several arguments supporting this scenario:
\begin{itemize}
\item Vigorous ongoing star formation without signs of quenching indicates ample reservoir of gas, while it is expected that a satellite galaxy would be stripped of its gas shortly after infall into the main halo. The star formation rate of the LMC has been unusually low until a recent burst starting $\sim 3$--4~Gyr ago \citep{Harris2009,Meschin2014,Hasselquist2021,Massana2022}, which might have been triggered by the compression of gas as it experiences a bow shock upon entering the Milky Way gas corona, although the interaction with the SMC is another possible explanation. \new{A lack of evidence for an earlier episode of elevated star formation rate (excluding the time shortly after the Big Bang) may be seen as the argument against a previous pericentre passage.}
\item Absence of a large-scale stellar tidal stream like that of the Sagittarius galaxy, which has completed several orbits around the Milky Way. Signs of tidal perturbation in the outer disc \citep{Mackey2016, Belokurov2019a, Cullinane2022} can be attributed to the interaction between LMC and SMC. There is a prominent gas stream \citep{Mathewson1974} that roughly matches the past orbit of the Magellanic Clouds over $>150^\circ$ on the sky; however, it is also better explained by the interaction between the two Clouds \citep{Besla2010,Diaz2012}.
\item LMC satellites, including the SMC, would have been tidally stripped from it if it had similarly close pericentre passage around the Milky Way in the past. 
\end{itemize}

\subsection{Past orbit}  \label{sec:past_orbit}

\begin{figure}
\includegraphics{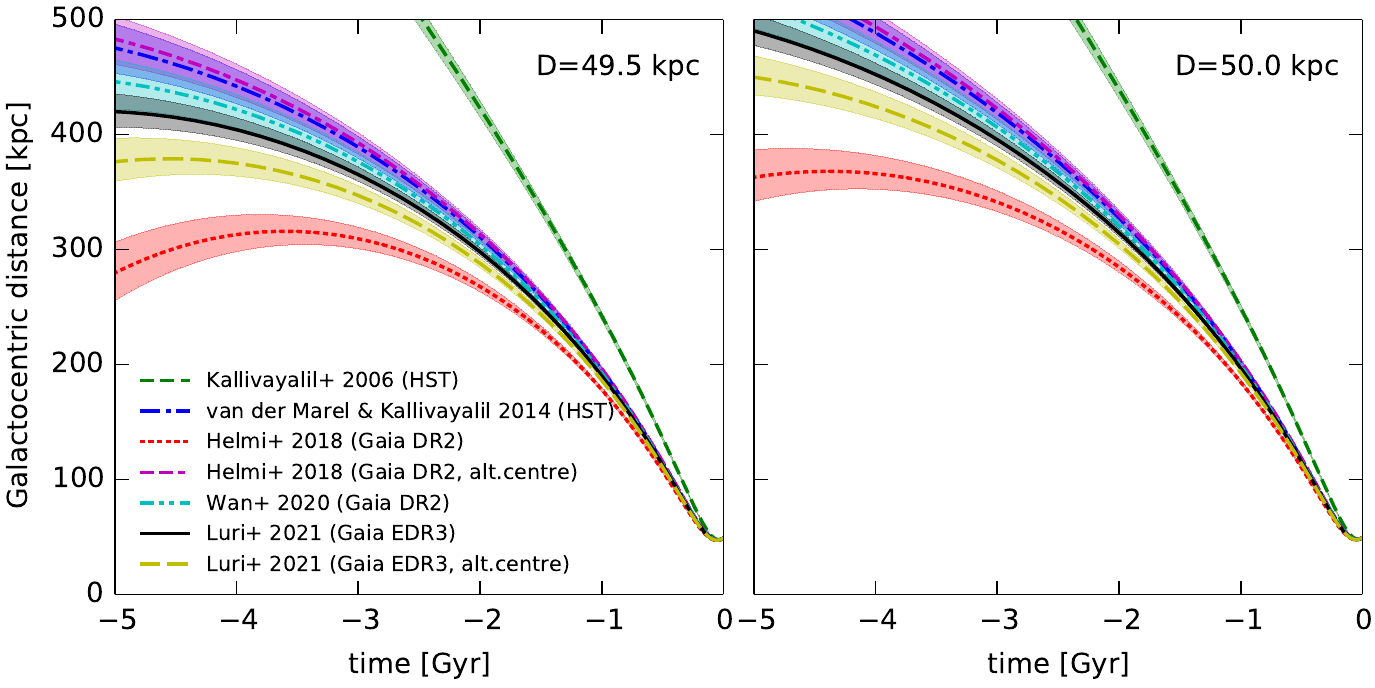}
\caption{Past trajectory of the LMC for various choices of the present-day position/velocity. Different lines correspond to sky positions and PM from different studies listed in Table~\ref{tab:LMCposvel}; shaded regions show the uncertainties associated with the PM and line-of-sight velocity ($262.2\pm 3.4$~\kms in all cases \citep{vanderMarel2002}). Left and right panels show the effect of changing the current distance by 1\%, which is comparable to the difference in PM between studies and far exceeds the nominal PM uncertainty in each study. The LMC follows a test-particle orbit in the fiducial potential described in the text.
}  \label{fig:trajectory_testparticle}
\end{figure}

The reconstruction of the past orbit of the LMC is a challenging task. If its mass were negligible, one could simply integrate the equations of motion backward in time for any assumed Milky Way potential. Even in this oversimplified case, the orbit parameters (apocentre distance and period) are surprisingly sensitive to minor variations in the assumed present-day phase-space coordinates. Figure~\ref{fig:trajectory_testparticle} shows the orbits computed in a fiducial Milky Way potential for several choices of the LMC PM and associated centre locations from Table~\ref{tab:LMCposvel}. Although the difference in 3d velocity between these studies (with the exception of the older \textit{HST} measurement from \citet{Kallivayalil2006a}) is only $\sim 10$~\kms, the orbital periods vary by a factor of two because the 3d velocity is quite close to the local escape velocity. A comparison between left and right panels shows that even a 1\% variation in the assumed distance changes the orbits very significantly; this change is driven by a $\sim 5$~\kms increase in the tangential velocity (same PM multiplied by a slightly larger distance) rather than a change in the position per se. This complicates the comparison of studies that adopt slightly different present-day positions and PM of the LMC without much discussion of the impact of these choices.

\begin{figure}
\includegraphics{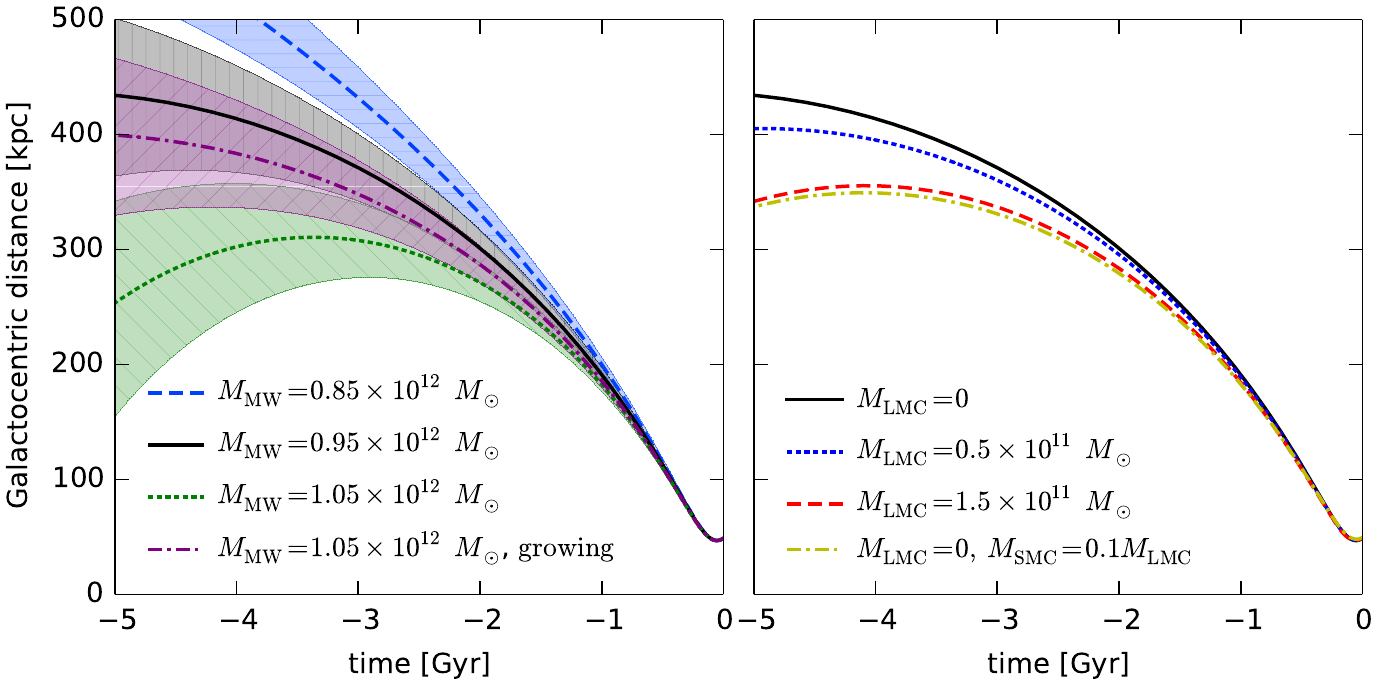}
\caption{Past trajectory of the LMC for different choices of the Milky Way potential (left panel) or LMC mass (right panel). 
The black curve corresponds to the fiducial potential with $M_{\rm MW}=0.95\times10^{12}\,M_\odot$, $M_{\rm LMC}=0$, present-day position and PM taken from \citet{Luri2021}, and heliocentric distance of 49.6~kpc \citep{Pietrzynski2019}. Blue and green curves in the left panel are plotted for a slightly lighter ($0.85\times10^{12}\,M_\odot$) or heavier ($1.05\times10^{12}\,M_\odot$) Milky Way, and the magenta curve represents the case when the Milky Way mass grows linearly over the Hubble time. Shaded regions show the 1$\sigma$ range of orbits arising from the uncertainty in the present-day position/velocity (chiefly the distance with a 1\% error bar). In the right panel, dotted blue and dashed red curves show the trajectory of a massive LMC ($0.5\times10^{11}\,M_\odot$ and $0.5\times10^{11}\,M_\odot$) in live $N$-body simulations correspond to the same Galactic potential and present-day phase-space coordinates as the black (test-particle) orbit. The dot-dashed yellow curve shows the orbit of a massless LMC, but with present-day coordinates corresponding to the centre of mass of the LMC--SMC system with a mass ratio of 10:1.
}  \label{fig:trajectory_potential_var}
\end{figure}

The LMC orbit is also very sensitive to the assumed Galactic potential. Figure~\ref{fig:trajectory_potential_var}, left panel, shows the evolution of Galactocentric distance for several choices of the potential. The adopted Milky Way density profile consists of three components: \\
\makebox[12mm][l]{bulge:} $\rho_{\rm bulge} \propto (1+r/0.2\,\mbox{kpc})^{-1.8}\,\exp\big[-(r/1.8\,\mbox{kpc})^2\big], \quad M_{\rm bulge}=1.2\times10^{10}\,M_\odot$; \\
\makebox[12mm][l]{disc:} $\rho_{\rm disc} \propto \exp\big[-R/3\,\mbox{kpc} - |z|/0.3\,\mbox{kpc}\big], \quad M_{\rm disc}=5\times10^{10}\,M_\odot$; \\
\makebox[12mm][l]{halo:} $\rho_{\rm halo} \propto r^{-1}\,(r+r_{\rm halo})^{-2}\, \exp\big[-(r/300\,\mbox{kpc})^4\big]$.\\
The halo density normalization is chosen to produce a circular velocity of 233~\kms at $R=8$~kpc. The virial mass of this model (referred to as $M_{\rm MW}$) is defined as the mass enclosed in a radius $r_{\rm vir}$ such that the mean density in this volume is 100 times the critical density of the Universe; $r_{\rm vir} = (M_{\rm MW}/10^{12}\,M_\odot)^{1/3}\times 260$~kpc. For the fiducial model, we set $r_{\rm halo}=14$~kpc, corresponding to $M_{\rm MW}=0.95\times10^{12}\,M_\odot$ and $M(<50\,\mbox{kpc})=0.41\times10^{12}\,M_\odot$. In this case, a test-particle orbit has a period of 11~Gyr. A mere 10\% variation in the Milky Way mass \new{(well within the current uncertainties)} changes the period to 30~Gyr or 6~Gyr.

\begin{figure}
\includegraphics{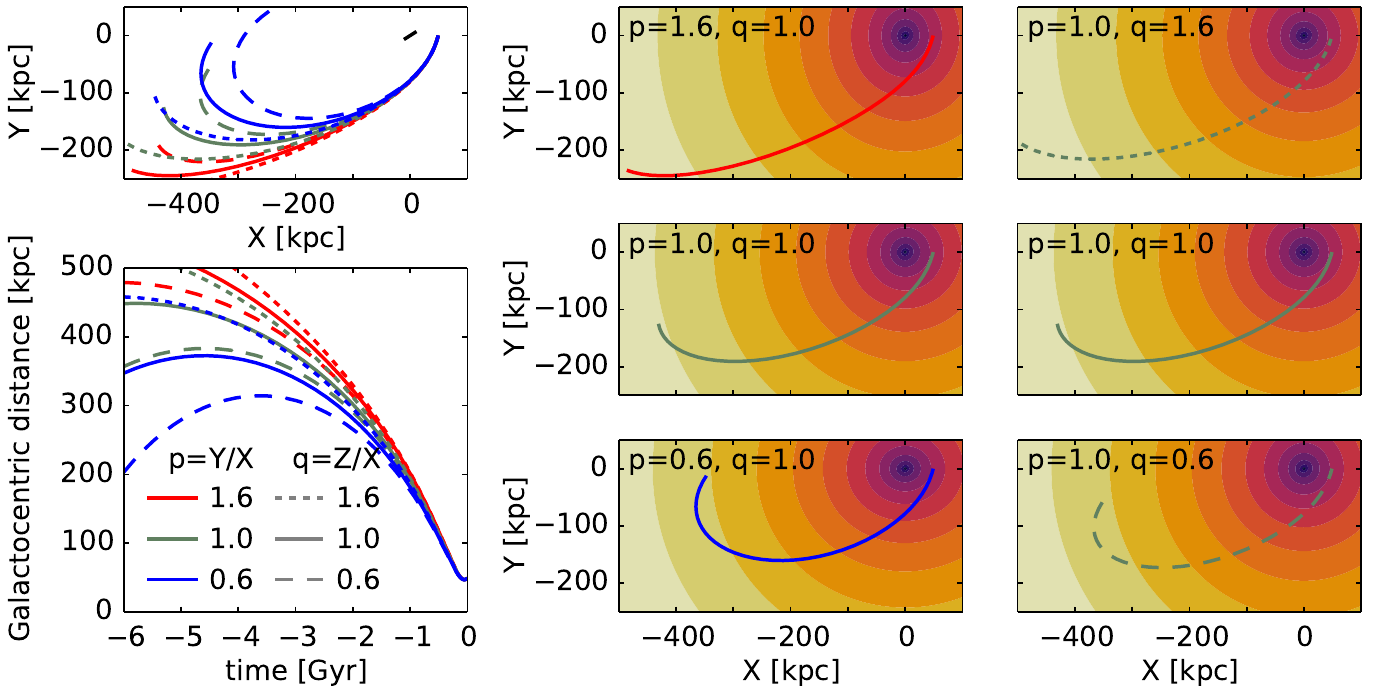}
\caption{Past trajectory of the LMC for different choices of the Milky Way halo shape parametrised by the axis ratios $p\equiv Y/X$ and $q\equiv Z/X$, where the orientation of the coordinate system is such that the LMC is currently located on the $X$ axis and its angular momentum points in the $Z$ direction. The Galactic potential has the same spherically-averaged mass profile in all cases, with the total mass $M_{\rm MW}=0.95\times10^{12}\,M_\odot$, and the LMC follows a test-particle orbit; the spherically symmetric case is identical to the one shown by the black curve in Figure~\ref{fig:trajectory_potential_var}.
The trajectory plane $X$--$Y$ plane is approximately perpendicular to the Galactic disc, which is shown by a black dash in the top left panel. The orbits for 9 different combinations of $p$ and $q$ are shown by different colours and dashes. Centre and right columns show the equipotential contours for different choices of $p$ and $q$, and the corresponding orbits.
The strong variation of orbital period and the apocentre radius is caused by the changes in the value of the potential at the current location of the LMC. This is more apparent for the centre column, where making the halo more flattened in the $Y$ direction (from top to bottom) stretches the equipotential surfaces horizontally, thereby lowering the value of the potential on the $X$ axis. On the other hand, increasing the flattening in the $Z$ direction (right column, from top to bottom) leaves the equipotential contours nearly round, but lowers the potential in the entire $X$--$Y$ plane. 
}  \label{fig:trajectory_mwshape}
\end{figure}

\new{As already mentioned, the shape of the Galactic halo is poorly known, and it also has a significant effect on the past orbit of the LMC. Figure~\ref{fig:trajectory_mwshape} shows test-particle orbits in several variants of a triaxial Milky Way halo, which all have the same spherically averaged density profile. Naturally, the halo shape and orientation affects the rate of precession of the orbit, but even more importantly, it also significantly changes the orbital energy and hence the period. To highlight this effect, we choose the orientation of the triaxial halo such that the $X$ axis points towards the current location of the LMC, and the $Z$ axis is parallel to its angular momentum vector. A halo that is flattened in the $Y$ direction has equipotential surfaces elongated along the $X$ axis, thereby lowering the value of the potential at the location of the LMC and making it more tightly bound. Flattening in the $Z$ direction has a similar but weaker effect, because the potential has lower values in the entire $X$--$Y$ plane compared to the value along the $Z$ axis at the same radius. The surprising strength of this effect is again due to the LMC being close to its pericentre, so that even small variations of the potential ($\sim 10\%$ between the most extreme cases in this example) significantly change the total energy (by a factor of two). Interestingly, the studies of the Orphan--Chenab \citep{Erkal2019,Koposov2023} and Sagittarius \citep{Vasiliev2021} streams, although disagreening in detail, generally favour an oblate halo shape with a minor axis roughly in the direction of the LMC angular momentum. A prolate halo with a major axis pointing towards the LMC is also a viable solution for the Orphan--Chenab stream. In both cases, the non-spherical shape of the halo effectively deepens the potential at the current location of the LMC, making its energy more negative and shortening its period.
}

The above analysis considered test-particle orbits of the LMC. However, a sufficiently massive LMC necessarily affects the orbits of stars in its vicinity. The most well-known manifestation of this influence is the accumulation of stars behind the moving object and the associated dynamical friction force created by this density wake. The classical \citet{Chandrasekhar1943} theory has been developed for the idealised case of motion on a straight line in an infinite homogeneous background, which is obviously never satisfied in real circumstances; nevertheless, it provides a reasonable order-of-magnitude estimate of the friction force and the magnitude of the overdensity in the wake. As is well known, the long-range nature of gravitational interactions leads to a logarithmic divergence of integral over the impact parameter of two-body interactions, which determines the overall friction force. In reality, the upper limit of this integral is imposed by the inhomogeneity of the background distribution, and the lower limit is set either by the distance of strong scattering ($\gtrsim 90^\circ$ deflection angle), or the characteristic size of the moving object, whichever is larger. The logarithm of the ratio of these two scales, known as the Coulomb logarithm $\ln\Lambda$, is customarily treated as a free parameter in the Chandrasekhar formula, calibrated against $N$-body simulations on account of approximate nature of the theory. In the simplest case, one would keep $\ln\Lambda$ constant; however, this approximation is rather crude, and a more refined version would make $\ln\Lambda$ depend on the orbital phase, usually by equating the maximum impact parameter with the instantaneous orbit radius
\citep{Tremaine1976,Hashimoto2003,Just2005}.

One may conclude that a more massive LMC needs to have a larger apocentre distance and higher initial energy, since it would have lost more energy to dynamical friction by present time. This argument led \citet{Kallivayalil2013} to the conclusion that a high mass of the LMC provides a stronger support for the first-infall scenario (see also Figure~1 in \citet{Besla2015}). However, this ignores another equally important factor, namely the reflex motion of the Milky Way in response to the LMC's gravitational pull. If both galaxies were point masses moving on Keplerian orbits about the common centre of mass, then increasing the mass of the LMC would shorten its period. This counteracts the effect of dynamical friction, but it is not obvious which of the two factors dominates. This question would be readily settled by $N$-body simulations, but the major problem is that for a reliable comparison of different runs, one needs to match the final (present-day) relative position and velocity of the LMC and the Milky Way to high precision. As already illustrated by Figure~\ref{fig:trajectory_testparticle}, even a 5~\kms difference in velocity can dramatically alter the past orbit. 

A popular technique for reconstruction of past orbits of \textit{both} galaxies (Milky Way and LMC) under mutual gravitational tug is based on the following approximation \citep{Gomez2015,Jethwa2016,Erkal2019,Patel2020,Vasiliev2021,Koposov2023}. Each galaxy is represented by a static (non-evolving) analytic potential $\Phi_{\rm MW}$, $\Phi_{\rm LMC}$ centered at $\boldsymbol x_{\rm MW}$, $\boldsymbol x_{\rm LMC}$ respectively. The centres of both galaxies satisfy the equations of motion
\begin{eqnarray}
\ddot {\boldsymbol x}_{\rm LMC} &=& \nabla\Phi_{\rm MW}(\boldsymbol x_{\rm LMC}-\boldsymbol x_{\rm MW}) + \boldsymbol a_{\rm DF}\big( \rho_{\rm MW}(\boldsymbol x_{\rm LMC}-\boldsymbol x_{\rm MW}), \dot{\boldsymbol x}_{\rm LMC}-\dot{\boldsymbol x}_{\rm MW} \big),  \nonumber \\
\ddot {\boldsymbol x}_{\rm MW} &=& \nabla\Phi_{\rm LMC}(\boldsymbol x_{\rm MW}-\boldsymbol x_{\rm LMC}) ,  \label{eq:orbit_reconstruction}
\end{eqnarray}
where $\boldsymbol a_{\rm DF}$ is the dynamical friction acceleration acting upon the LMC (usually incorporating a distance-dependent Coulomb logarithm). This ODE system is integrated backward in time starting from the present-day position and velocity of the LMC $\boldsymbol x_{\rm LMC}, \dot{\boldsymbol x}_{\rm LMC}$ and setting those of the Milky Way to zero (in effect, using the inertial reference frame in which the Milky Way is currently at rest). Note that the reference frame tied to the Milky Way centre at all times is not inertial, which needs to be taken into account when computing trajectories of other objects in the combined gravitational field of the Milky Way and the moving LMC. These equations are manifestly asymmetric (the mutual forces do not satisfy Newton's third law) and ignore the tidal distortions of both galaxies, yet they account for the reflex motion, at least qualitatively. 

Using this approach, \citet{Gomez2015} argued that accounting for the reflex motion of the Milky Way significantly reduces the apocentre radius of the LMC compared to the case with only the dynamical friction taken into account. However, in their Figure~1, a higher-mass LMC still had a larger apocentre than a lower-mass (in effect, the dynamical friction dominated over the reflex motion). On the other hand, \new{more recent applications of this method by \citet{Patel2017,Patel2020}, as well as full $N$-body simulations,} seem to indicate that the reflex motion is actually more important, i.e., higher-mass LMC has smaller apocentre radius and shorter period (Figure~\ref{fig:trajectory_potential_var}, right panel). The simulations shown on that figure were conducted in a carefully controlled way, iteratively refining the initial conditions for the orbit, so that the present-day position/velocity of the LMC matches the observations to within a fraction of kpc and \kms, ensuring a fair comparison with the test-particle integrations. Note that this trend is opposite to the one shown in Figure~1 of \citet{Besla2015}, which ignores the reflex motion correction. Therefore, an LMC of mass (1--2)${}\times10^{11}\,M_\odot$ can be on the first infall only if the Milky Way mass is below $10^{12}\,M_\odot$, contrary to many previous studies. 

It should be noted that the reconstruction of past orbit of the LMC is usually performed assuming an isolated (non-growing) Milky Way; however, there are compelling reasons to believe that our Galaxy has substantially grown in mass over the last 10~Gyr. Naturally, this implies a less bound (or more unbound) orbit of the LMC in the past: \citet{Kallivayalil2013} find that the inclusion of a cosmologically motivated Milky Way mass evolution \new{(unfortunately, without providing a detailed recipe)} increases the LMC period by $\sim 20$--30\%, strengthening the case for the first infall. Magenta curve in the left panel of Figure~\ref{fig:trajectory_potential_var} shows the orbit in a very simple model of the Milky Way mass growing linearly with time \new{from zero at Big Bang to its present-day value}: this produces an effect comparable with a $10\%$ reduction of its mass at present moment.

As mentioned in Section~\ref{sec:SMC}, even a comparatively puny SMC may cause a noticeable offset in the LMC's position and velocity relative to their common centre of mass. Yellow curve in the right panel of Figure~\ref{fig:trajectory_potential_var} demonstrates that a mass ratio of 10:1 produces a reduction of the orbital period comparable to the change of the LMC mass from 0 to $1.5\times10^{11}\,M_\odot$ -- in other words, a non-negligible effect that should be taken into account in future studies. \new{\citet{Patel2020} found that including the SMC in their analysis led to minor and occasionally moderate changes in the orbits of other Milky Way satellites, but did not discuss how it affects the inferred orbit of the LMC.}

In summary, the past orbit of the LMC is very sensitive to various seemingly minor changes in the model assumptions and the current coordinates. A more massive LMC and a higher mass ratio between the SMC and the LMC both cause the orbital period to decrease relative to the massless case, whereas a cosmologically motivated growth of the Milky Way mass moderately increases the period. A detailed study taking into account all these factors still has to be performed, and the $N$-body simulations would be the most realistic approach, but they are hampered by the need for extreme precision in matching the present-day coordinates of all galaxies to ensure a fair comparison between models. We conjecture that the first-approach scenario remains viable, but only if the Milky Way mass is below $10^{12}\,M_\odot$. 
It is interesting to note that a low mass of both the Milky Way and the LMC is advocated by \citet{Hammer2015,Wang2019} based on their simulations of the Magellanic stream, though this view is not shared by the majority of current studies.
An intriguing alternative would be that both galaxies are more massive than usually assumed, therefore a previous pericentre passage would have occurred a few Gyr ago, but at a much larger distance ($\sim100$~kpc) than the most recent one, especially when accounting for the time evolution of the Milky Way mass. In this case, the tidal radius of the LMC at this point would remain large enough ($30-40$~kpc) to retain most of its mass and satellites, negating one of the arguments in favour of the first passage. Obviously, a reliable simulation of this scenario is even more challenging \citep[e.g.,][]{Guglielmo2014}.

\section{Dynamical implications}  \label{sec:dynamics}

Having established that the LMC is likely rather massive and just passed its first pericentre around the Milky Way, we now consider the implications of this scenario, which range from localised perturbations to far-reaching consequences within the entire Local Group.

\subsection{Local effects}  \label{sec:local_effects}

The most obvious effect of a massive moving object is the perturbation of stars in its immediate vicinity. As discussed above, it creates an overdensity of stars behind the moving object (``wake''), which in turn is responsible for dynamical friction. \citet{Belokurov2019b} provided observational evidence for such a wake in the form of the so-called Pisces plume, detected both as an overdensity in the 3d distribution of RR Lyrae stars and as a kinematical offset in the co-spatial population of BHB stars, qualitatively consistent with \new{the wake signatures predicted  by \citet{GaravitoCamargo2019} from $N$-body simulations}.

A close passage of the LMC deflects the orbits of several Milky Way satellites and other objects in the Galactic halo. \citet{Simon2020} determined that a recently discovered ultrafaint dwarf galaxy Tucana~IV approached to within a few kpc from the LMC $\sim120$~Myr ago, and possibly had a comparably close encounter with the SMC, but in both cases with a high enough relative velocity that a bound orbit about either Cloud is unlikely. Nevertheless, its orbit around the Milky Way had been significantly altered as a result of this interaction. A few other galaxies have also passed within 20--30~kpc from the LMC in the last few hundred Myr, including Aquarius~II, Grus~II, Sagittarius~II, Tucana~III, Tucana~V, but again with a high enough velocity to exclude long-term dynamical association (see Figure~9 in \citep{Battaglia2022} and Figure~12 in \citep{CorreaMagnus2022}). This list does not include the likely satellites of the LMC discussed in Section~\ref{sec:lmc_satellites}, nor the other galaxies whose orbits have been indirectly affected by the LMC through the induced reflex motion of the Milky Way, as discussed later in Section~\ref{sec:local_group}.

Last but not least, a close encounter with the LMC may have a significant effect on thin stellar streams. In a static Milky Way potential, stars in a tidal stream follow very similar (though not exactly the same) orbits \citep{Eyre2011,Sanders2013}, and their velocity vectors are expected to be aligned with the stream track (even in a non-spherical potential). In practice, we do not measure the 3d velocities directly, but only their sky-plane components from PM, and for a subset of stars with spectroscopic follow-up observations, the line-of-sight component. Both have contributions from the Solar velocity, which needs to be subtracted -- this is straightforward for the line-of-sight component, but compensating the contribution of Solar reflex to the PM requires an accurate estimate of the distance. Assuming that this can be done, the next step is to compare the reflex-corrected values with the gradient of the distance to the stream (for the line-of-sight component) and the stream track on the sky (for the PM); the former is usually much less well known. Any misalignment could be taken as an indication of some form of non-stationarity in the gravitational potential -- be it a close passage of the LMC (or another massive object such as the Sagittarius dwarf galaxy \citep{Dillamore2022}), acceleration and deformation of the Milky Way induced by the LMC \citep{Vasiliev2021,Lilleengen2023}, or a figure rotation of the potential \citep{Valluri2021}. 

The above mentioned Tucana~III galaxy has a highly eccentric Galactocentric orbit with a pericentre distance of only a few kpc, and thus undergoes tidal disruption producing an observed stream. \citet{Erkal2018} conjectured that the LMC may induce a PM component in the direction perpendicular to the stream, and predicted that it should be detectable in the upcoming \textit{Gaia} DR2 catalogue. However, a subsequent analysis of \textit{Gaia} data by \citet{Shipp2019} did not corroborate this prediction, which they attributed to a possible systematic offset in the distance to the stream (which is crucial for the Solar reflex correction) or due to oversimplifications of the model.
On the other hand, such a misalignment between the stream track and the reflex-corrected PM vectors was clearly detected by \citet{Koposov2019} and \citet{Fardal2019} for the very long Orphan--Chenab stream (which originally was discovered separately in the Northern and the Southern hemispheres, hence the dual name). \citet{Erkal2019} demonstrated that it can be well explained by a perturbation from the LMC if its mass is $(1.3\pm0.3)\times10^{11}\,M_\odot$, simultaneously constraining the Milky Way mass within 50~kpc with a 10\% uncertainty. In this model, the stream was deflected by the LMC passing within 20--30 kpc from the southern portion of the stream.
\citet{Shipp2021} extended this analysis to a few other streams in the Southern hemisphere, finding that four of them (including Orphan--Chenab) produce a consistent estimate of the LMC mass (although they kept the Milky Way potential fixed). The final value from all four streams combined was $M_{\rm LMC}=(1.8\pm0.2)\times10^{11}\,M_\odot$, and given that the distance of closest approach is different for each stream, such a combined analysis opens up the possibility of constraining not just the total mass, but the mass profile of the LMC.
\citet{Koposov2023} repeated the fit of the Orphan--Chenab stream with a more flexible model and using updated astrometric and spectroscopic observational data. They obtained the total LMC mass of (1--1.5)${}\times10^{11}\,M_\odot$, constrained its mass profile (most tightly around 30~kpc, with a 13\% uncertainty) and provided even tighter constraints on the Milky Way mass profile and shape in the radial range 20--50~kpc, with the smallest relative uncertainty of only 5\% around 30~kpc. Thus the emerging consensus is that a massive LMC with \mbox{$M_{\rm LMC}=(1$--$2)\times10^{11}\,M_\odot$} is needed to explain the deflection of several streams.

Another interesting ``local effect'' of the LMC is not dynamical, but rather has implications for direct dark matter detection experiments. Namely, the dark halo of the LMC itself likely extends beyond the 50~kpc distance to its centre, and therefore we may expect to find dark matter particles from the LMC in the Solar neighbourhood. Crucially, these particles have very high velocity relative to the Sun ($\sim 500$--$800$~\kms), and even if their density is tiny compared to the Milky Way halo itself, the high velocities can produce a prominent signal in direct detection experiments -- or rather, the absence of detection places stronger constraints on the properties of dark matter particles \citep{Besla2019,Donaldson2022,SmithOrlik2023}. 

\subsection{Global effects on the Milky Way}  \label{sec:global_effects}

\subsubsection{Theory}  \label{sec:global_effects_theory}

\begin{figure}
\includegraphics[width=14cm]{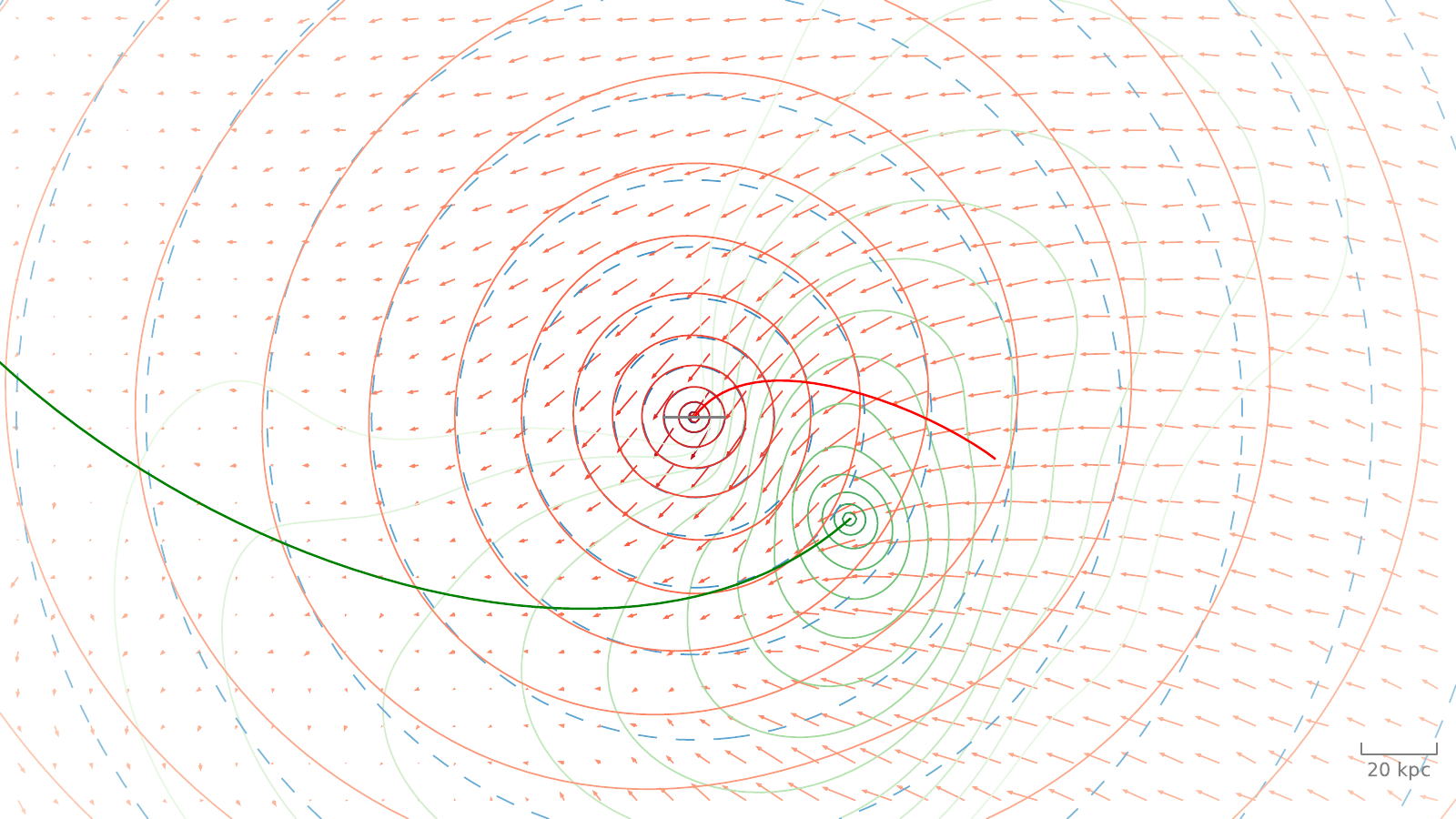}
\includegraphics[width=14cm]{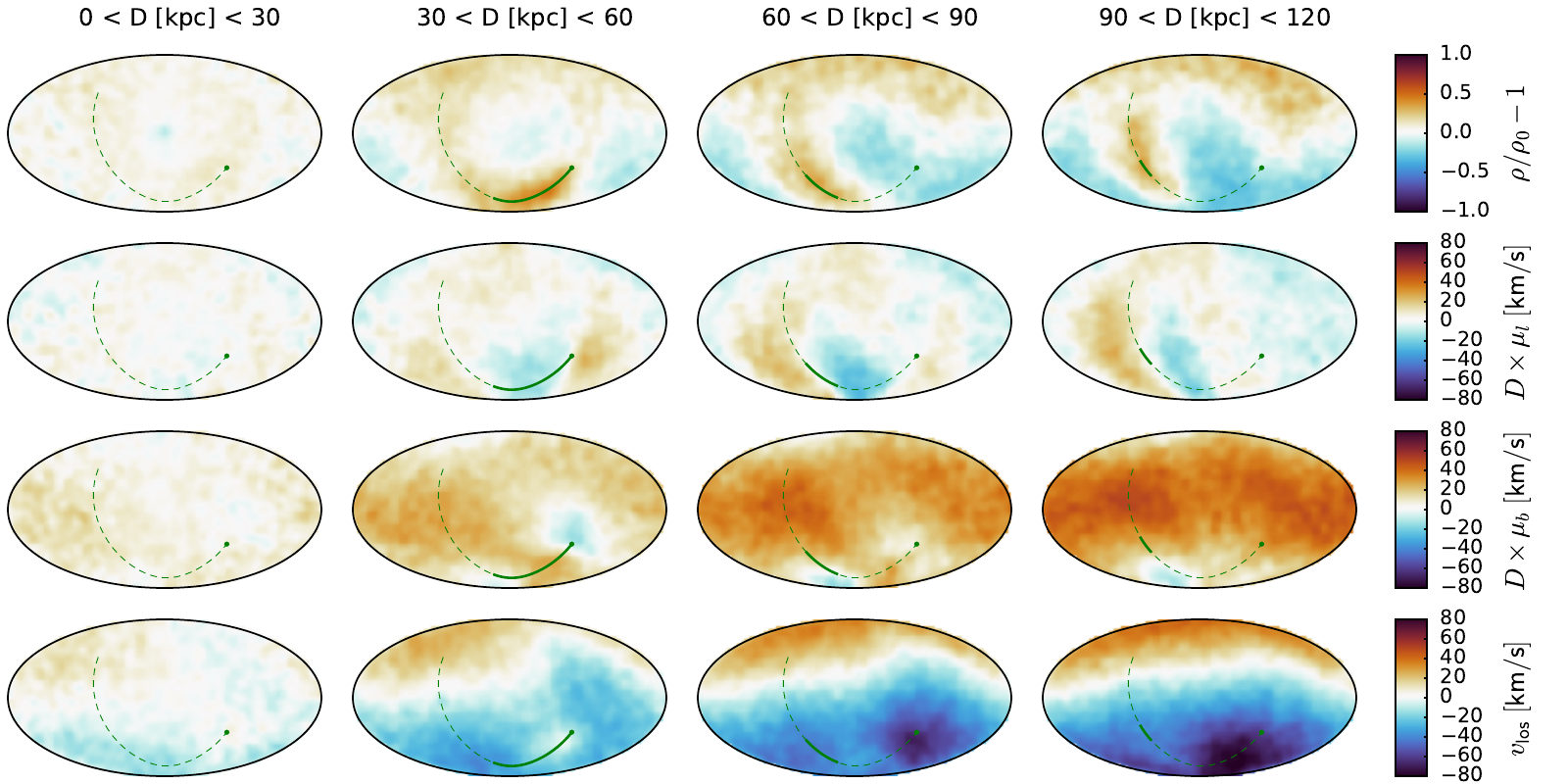}
\caption{\textit{Top panel:} side view of the final snapshot in the $N$-body simulation of a Milky Way--LMC interaction. The Galactic disc is oriented edge-on (gray line), and the Sun's position is directly in front of the image plane, which is close to the orbital plane of the LMC. The past trajectories of both galaxies are shown by red and green curves. Red and green contours show the projected density of the Milky Way and LMC haloes, respectively, while the dashed blue contours show the initial Milky Way halo (spherical). Regions where the red contours are closer to or further away from the Galactic centre than the blue ones appear as under- or overdensities, respectively. Arrows display the mean velocity of Milky Way particles in the centre-of-mass frame, which reaches 60~\kms in the central region. In the non-inertial reference frame associated with the Galactic centre, the mean velocity in the central 20-30 kpc is close to zero, but in the outer regions it is primarily directed upward (one needs to subtract the central velocity vector from all vectors shown in the plot). \protect\\
\textit{Bottom panel:} the same snapshot as viewed from the Solar location at rest w.r.t.\ the Galactic centre (i.e., corrected for the Solar motion). Shown are sky maps of over/under density (top row), mean velocity in the $l$ and $b$ directions (PM multiplied by distance, second and third rows), and mean line-of-sight velocity (bottom panel). The snapshot is split into four bins in heliocentric distance, shown in columns from left to right. The innermost one has very little perturbation in either density or kinematics, but it becomes progressively larger further out. Apart from localized overdensities along the past LMC orbit, the dominant feature is the north--south dipole asymmetry in density and line-of-sight kinematics, and the positive mean $v_b$ (upward motion in the sky plane) in the outer Galaxy.
}  \label{fig:model}
\end{figure}

The deflection of nearby stars is only part of the story, however. A more subtle but equally important effect is associated with the reflex motion of the Milky Way induced by the LMC and the associated distortions in the Galactic halo. Figure~\ref{fig:model} illustrates this phenomenon, using the present-day snapshot from the fiducial $N$-body simulation with $M_{\rm MW}=0.95\times10^{12}\,M_\odot$ and $M_{\rm LMC}=1.5\times10^{11}\,M_\odot$; the effect remains qualitatively the same in different Milky Way potentials, but scales in proportion to the LMC mass.

In the inertial frame, both galaxies move around the common centre of mass. Top panel shows the edge-on view of the simulation. The LMC moves diagonally towards the upper right corner on this plot, and the Milky Way, correspondingly, in the opposite sense. However, the mean velocity of stars in the Galactic halo is not uniform in space: whereas the inner regions (few tens kpc) move with the same velocity as the Galactic centre, and therefore appear stationary in the Galactocentric reference frame, the outer regions have experienced different amounts of gravitational acceleration from the LMC. The region on the left, for instance, was first pulled towards the LMC as it was approaching from the left side of the plot, and then pulled in the opposite direction (towards the Galactic centre) as the LMC moved right; the net velocity change turns out to be close to zero. The region on the right, on the other hand, was steadily accelerated towards the LMC (and hence the Galactic centre) with a larger accumulated velocity change than the central region itself. But these horizontal motions pale in comparison to the vertical ($z$-component) velocity: the inner Milky Way rapidly swings towards the LMC, which passes below the Galactic disc (in the Southern hemisphere) in the last few hundred Myr, but the outer halo did not have time to react to this sudden change in velocity of the inner Galaxy, nor had it experienced a comparable amount of downward acceleration from the LMC itself (since it was more distant). As a result, the net velocity of the outer halo in the Galactocentric reference frame is directed upwards in $z$ and slightly inwards in $R$. 

A comparison of red (present-day) and blue (initial) density contours shows that this sudden downward displacement of the inner Galaxy w.r.t.\ the outer halo resulted in an apparent overdensity in the Northern hemisphere and underdensity in the South. The reason is again the same: stars at large distances towards the North Galactic pole have a smaller mean $z$-component of velocity than the inner Galaxy, so they have moved downward a smaller distance than the central region, and therefore currently find themselves at a larger Galactocentric radius than originally. Since the halo density drops with radius, these ``impostors'' from smaller initial radii retain a higher-than-expected density at their current location. The situation is opposite in the Southern hemisphere.

Bottom panel of the same figure shows how these perturbations should appear to an observer at the Solar location and at rest w.r.t.\ the Galactic centre. The inner 30 kpc do not show significant perturbation, since all stars move in sync with the origin of the reference frame itself. The asymmetries become progressively larger further out. In the density plot (top row), one can discern the local wake along the past trajectory of the LMC in each distance bin, superimposed with the global north--south dipole asymmetry. The net upward motion of the outer halo creates a similar dipole asymmetry in the line-of-sight velocities (bottom row) and a positive bias in the $b$ component of PM (third row), whereas the $l$ component (second row) largely shows the localized signal from the wake.

To summarize, the large-scale perturbation from the LMC appears because the Galaxy does not move as a rigid body in response to its gravitational pull. Although the dynamical consequences of the reflex motion of the Milky Way induced by the LMC were recognised long ago by \citet{Weinberg1995} and emphasised in \citet{Gomez2015}, the importance of \textit{differential} motion (i.e., distortion) was first pointed out by  \citet{Erkal2019}, who predicted a net upward motion in the outer halo. \citet{GaravitoCamargo2019,GaravitoCamargo2021a,Erkal2020b,Petersen2020,Cunningham2020} and \citet{Makarov2023} conducted simulations of the MW--LMC encounter and presented maps of observable signatures similar to the ones described above (note that the abscissa axis is flipped in some of these plots). Although the details of these simulations differ somewhat, the predicted perturbation features are qualitatively similar and not very sensitive to the particular choice of the Milky Way potential (unlike the entire past trajectory of the LMC), since they are mainly created in the last few hundred Myr as the LMC plunges below 100~kpc. 
By and large, the inner part of the Milky Way remains unperturbed; simulations conducted by \citet{Laporte2018a} indicated that the LMC passage induces a warp in the stellar and gas discs at distances $\sim 20$~kpc, but its amplitude ($\lesssim 1$~kpc) is significantly smaller than the observed one even for the highest mass LMC they considered ($2.5\times10^{11}\,M_\odot$).
All these simulations assume a first approach trajectory; the distortions in the Milky Way halo in the case of more than one pericentre passage would likely be different, but remain unexplored.

\subsubsection{Observations}  \label{sec:global_effects_observations}

\begin{figure}
\includegraphics{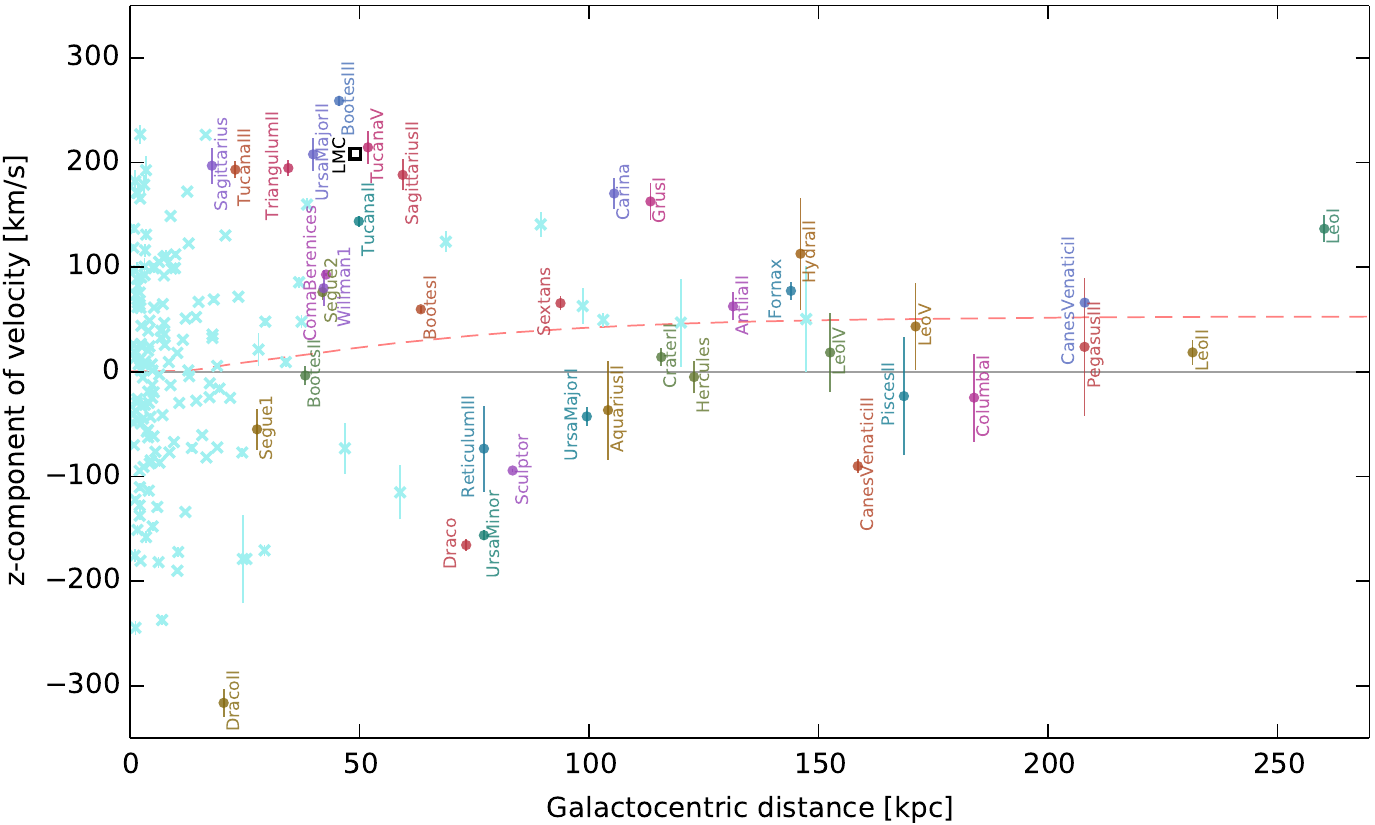}\\
\includegraphics{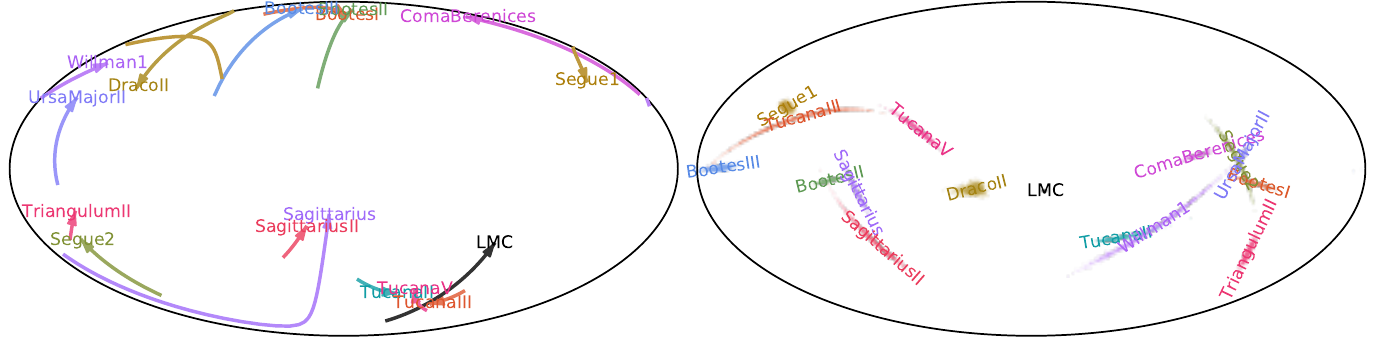}
\caption{\textit{Top panel:} $z$-component of Galactocentric velocity for Milky Way globular clusters (cyan crosses) and satellite galaxies (coloured circles; the colour only serves to distinguish them more easily) plotted against Galactocentric distance. The LMC is shown by a black square, and its satellites are excluded from the plot; nevertheless, there is a conspicuous concentration of a dozen unrelated galaxies with positive $v_z$ within 70 kpc. Further out, the bias towards positive $v_z$ is less prominent but still noticeable. Red dashed line shows the mean $v_z$ in the present-day snapshot of the fiducial $N$-body simulation shown in Figure~\ref{fig:model}. It does not explain the coherent group of galaxies with $v_z\simeq 200$~\kms in the inner region, but qualitatively matches the positive bias in the outer region. \protect\\
\textit{Bottom left panel:} orbits of selected satellites with current Galactocentric distances under 70~kpc, computed for the last 100~Myr and shown in Mollweide projection as viewed from the Galactic centre. Although many of them have similar values of $v_z\simeq 200$~\kms, they are not located in the same region of space (except the three Tucana and three Bo\"otes galaxies), nor share common orbits. \protect\\
\textit{Bottom right panel:} directions of orbital poles for the same selection of orbits (similar to the top panel of Figure~3 in \citep{Fritz2018} or the left panel of Figure~9 in \citep{CorreaMagnus2022}). Again, there is no significant non-uniformity in this distribution, suggesting that the velocity values are purely coincidental.
}  \label{fig:meanvz}
\end{figure}

These rather unambiguous and prominent predicted signatures have been supported by observational evidence. As soon as accurate measurements of 3d velocities of globular clusters and galactic satellites became available (after \textit{Gaia} DR2), it was noticed that there is a positive mean $v_z$ in the population of satellites beyond 30 kpc \citep{Erkal2020b,Petersen2021}. Figure~\ref{fig:meanvz}, top panel, demonstrates that a handful of globular clusters and most of the 20-odd satellite galaxies beyond 80 kpc have an average $v_z$ of a few tens \kms, consistent with expected signature from a massive LMC. On the other hand, the largest asymmetry in the $z$-velocity of satellites is observed at smaller distances (20--60~kpc), where many galaxies have $v_z\sim 200$~\kms -- including the LMC itself, but excluding objects even tentatively marked as its satellites.  However conspicuous this concentration might be, it seems to be largely accidental, since these objects are neither close in space nor share a common orbital plane (bottom panel). Given this bizarre coincidence, which cannot be caused by the LMC-induced net velocity, we might just as well disregard the weaker positive bias further out in the halo, but it certainly doesn't contradict the model.

The bias towards positive $v_z$ should also be manifested in the line-of-sight velocity as a dipole north--south asymmetry (bottom row in Figure~\ref{fig:model}). Its detection has been independently announced by \citet{Erkal2021} and \citet{Petersen2021}, using mostly the same spectroscopic sample of BHB stars and K giants from the SEGUE survey \citep{Xue2011,Xue2014} (for these types of stars, the distance can be measured with a $\sim10\%$ precision, though in \citep{Erkal2021} it was only used to restrict the sample to the outer halo, and not for the analysis of transverse velocities due to relatively large PM errors). Ref.\ \citep{Petersen2021} also determined the orientation of the velocity dipole (the apex vector, in their terminology) to be significantly tilted (by $\gtrsim 55^\circ$) from the South pole towards the direction where the LMC was a few hundred Myr ago, see their Figure~1. By contrast, \citet{Erkal2021} only examine the north--south gradient in the line-of-sight velocity. The models generally predict that the dipole points towards a more recent location of the LMC, some 15--20$^\circ$ in the opposite direction from the South Galactic pole than in \citep{Petersen2021}. At the moment, the reason for this tension is unclear and not much discussed in the literature.

The signal in the $b$-component of the PM is more difficult to detect, given relatively large measurement uncertainties ($\sim 0.2$~mas\,yr$^{-1}$ for an RR Lyrae at a distance of 50~kpc, corresponding to the velocity uncertainty of 100~\kms) and the need to accurately measure the distance in order to subtract the solar reflex velocity. Currently there is no conclusive evidence for such a signal. It should be noted that a net rotation of the stellar halo could also manifest itself in large-scale asymmetries in the line-of-sight velocity and PM maps, but they have different decomposition into multipoles than the LMC-induced linear motion. On the other hand, unmixed debris from various accreted and tidally disrupted galaxies may confound the signal encoded in the smooth halo \citep{Cunningham2020}; in fact, the removal of the largest of these -- the Sagittarius stream -- is a perennial problem in the studies of halo structure and kinematics.

The density asymmetries induced by the LMC (or any other features in the density profile, for that matter) are generally more difficult to measure with certainty. Unlike kinematic patterns, the inference about the density profile is much more dependent on the precise knowledge of the selection function of the survey (the probability of an actually existing star to enter the catalogue), and the latter depends on many factors that are often difficult to quantify and control. Even an all-sky survey such as \textit{Gaia} does not have a uniform sensitivity and coverage due to the way it scans the sky \citep{Boubert2020b,CantatGaudin2023}, and signatures of this scanning law are apparent in various star count maps at faint magnitudes. These complications notwithstanding, \citet{Conroy2021} used \textit{Gaia} (optical) and WISE (infrared) photometry to select a sample of red giant stars beyond 60~kpc, which was found to display a prominent asymmetry pattern similar to the one shown in the right two columns of top row of Figure~\ref{fig:model}. They interpreted it as a signature of the LMC influence -- both local (the wake along its past trajectory in the south, also detected in \citep{Belokurov2019b} as described earlier) and global (offset of the outer halo creating an overdensity in the north, which they call ``collective response''). However, the amplitude of the observed asymmetries was $\sim2\times$ higher than even the heaviest-LMC model predicted. On the other hand, in \citet{Chandra2023} the same team analysed their spectroscopic sample of distant halo stars in the same regions as these overdensities, and conjectured that they can be attributed to debris from an early massive merger in the Milky Way history, which remain on highly radial orbits tilted at $\sim 45^\circ$ from the Galactic disc.

\new{It is worth emphasising that in this review we used the term ``reflex motion'' in the purely mechanical sense, as the displacement and associated motion of our Galaxy about the common centre of mass of the Milky Way--LMC system (analogous to the wobbling of a star induced by its planets). However, this definition is impractical because this centre of mass is not an observationally accessible concept, and moreover the Galactic centre does not coincide with the centre of mass of the entire Milky Way. Other studies \citep{GaravitoCamargo2021a,Petersen2021} use a more pragmatic approach and define the reflex motion as the velocity of the Milky Way centre w.r.t\ the outer parts of its halo, implicitly assuming that the latter stays fixed. As the red dashed line in Figure~\ref{fig:meanvz} shows, this velocity difference reaches an asymptotic value beyond $\sim 100$~kpc, making that region a good proxy for the inertial frame; on the other hand, the number of suitable kinematic tracers drops rapidly beyond 50~kpc. Likewise, the distortions of the Milky Way halo shape resulting from the accumulation of kinematic differences with time have a variety of names in the literature. \citep{Weinberg1995,GaravitoCamargo2021a,Conroy2021} use the term ``collective response'', but unlike collective modes and instabilities in plasma and self-gravitating systems, self-gravity plays no role in this phenomenon \citep{Rozier2022} -- stars respond similarly not because they interact with each other, but because the potential in which they move changes in the same way. }

\subsubsection{Implications for the Milky Way dynamics}  \label{sec:global_effects_implications}

The global upward motion of the outer halo w.r.t.\ the Milky Way centre \new{affects the dynamics of various components of our Galaxy, for instance, stellar streams.} The largest and most complex of these -- the Sagittarius stream -- stretches across the entire great circle and spans a range of Galactocentric distances from 15 to more than 100~kpc. Its 3d track, especially around the apocentre of the leading arm, is difficult to explain: a widely used model from \citet{Law2010} invoked a rather unusual shape of the Milky Way potential, with its minor axis lying in the Galactic plane and roughly aligned with the orbital pole of the LMC (which is also on a nearly polar orbit, but orthogonal to that of the Sagittarius). They already conjectured that this might be an artifact of neglecting the gravitational effect of a massive LMC -- a suspicion reinforced by \citet{VeraCiro2013} and \citet{Gomez2015}. However, neither of these studies considered the consequences of the Milky Way deformation, nor they had access to full 6d phase-space coordinates of the stream stars. After these became available in \textit{Gaia} DR2, \citet{Vasiliev2021} discovered that the reflex-corrected PM vectors are misaligned with the stream track, like in the Orphan--Chenab stream \citep{Koposov2019}. A comprehensive modelling effort performed in \citep{Vasiliev2021} demonstrated that this misalignment, along with the shape of the leading arm, can only be explained by invoking a time-dependent perturbation of a massive LMC with $M_{\rm LMC}=(1.3\pm0.3)\times10^{11}\,M_\odot$. Critically, the main effect of the LMC is not a local deflection of the stream, as in the case of Orphan--Chenab, but the downward reflex motion of the Milky Way. Figures~7 and 12 in that paper demonstrate that when the Milky Way potential is artificially pinned down at origin, the shape of the stream is completely wrong and the longitudinal variations of energy and angular momentum along the stream even reverse sign.

Another example of a physical situation where the reflex motion is crucial to take into account is the deflection of hypervelocity stars \citep{Boubert2020a}. Figure~1 in that paper shows that such a star ejected from the Galactic centre on a nearly radial trajectory would be deflected downward by the LMC if the Milky Way were not moving, but would appear to be deflected upward (like the rest of the outer halo) when accounting for the reflex motion.

In the above examples, the displacement (reflex motion) of the Milky Way left prominent signatures in the spatial distribution and kinematic properties of \textit{tracer populations} (halo stars, streams, satellites). However, as the dark halo of the Galaxy would be subjected to the same deformations, the gravitational potential of the Milky Way is also distorted, which would create further perturbations in the tracer orbits. \citet{Lilleengen2023} examined the importance of these next-order perturbations for the modelling of the Orphan--Chenab stream, using a basis-set representation of potentials of both galaxies. This technique allows one to selectively enable or disable different terms in the spherical-harmonic expansion describing the deformation. First a real $N$-body simulation of the Milky Way--LMC interaction is performed with all deformations inherently present. Then a simulated stream is evolved in the pre-recorded time-dependent potential of interacting galaxies under different assumptions about their deformations. They find that the dominant contribution comes from the Milky Way dipole (i.e., upward displacement of the outer halo), followed by monopole and quadrupole perturbations in the LMC (tidal shock and formation of two tails), and that neglecting these factors has a measurable effect on the thin and cold stream. On the other hand, a similar analysis performed in \citep{Vasiliev2021} for the Sagittarius stream concluded that these deformations are relatively unimportant, presumably due to this stream being much broader and hotter. \new{In any case, the distortions of the Milky Way halo caused by the LMC undoubtedly complicate the analysis of its shape, which so far has been inconclusive.}

Using a different technique (matrix method from kinetic theory), \citet{Rozier2022} found that accounting for the self-gravity of the Milky Way halo (i.e. changes in the potential resulting from the distortions) does not significantly affect the amplitude of its response to the LMC pull. The global dipole perturbation (essentially, the differential reflex motion) is also rather insensitive to the kinematic structure of the halo (velocity anisotropy). On the other hand, the local wake is relatively stronger in radially biased models than in tangentially biased ones, and its geometry also varies with anisotropy. \new{By contrast, \citet{Makarov2023} find that particles on near-circular orbits at radii 50--100~kpc exhibit a larger kinematic offset from the inner Milky Way than particles with low angular momentum in the same radial range.}

Given the substantial perturbations (on the order of few tens percent) in the density and kinematics of the Milky Way halo, one may question the applicability of classical dynamical modelling methods for determining the Galactic gravitational potential, which are almost always based on the equilibrium assumption. \citet{Erkal2020b} found that neglecting these perturbations biases up the inferred mass by 15--20\% beyond 50~kpc (for an LMC mass of $1.5\times10^{11}\,M_\odot$). They used the tracer mass estimator from \citep{Watkins2010}, which approximates the potential and the tracer density as power laws, assumes a constant anisotropy, and invokes the spherical Jeans equation to link these quantities. Despite its simplicity, this method is widely used for measuring the mass profiles of the Milky Way and other galaxies. They also found that the bias is slightly reduced when using the velocity dispersion rather than the full second moment of velocity, and could be minimised by choosing a particular region of the halo least affected by velocity and density perturbations.

\citet{Deason2021} used a related class of power-law mass estimators, but relying on the distribution function rather than the Jeans equation, to measure the Milky Way mass profile out to 100~kpc from the kinematics of a few hundred stars in the outer halo (the sample from \citep{Erkal2021}). They tested the method on simulated data including the LMC perturbation, and found that it returns a value biased up by $\lesssim 10\%$; this bias is lower than in the previous study in part due to a fortunate location of the observational sample in the sky regions less affected by the LMC. In fact, biases from unrelaxed structures in the halo remaining from other accretion events (in particular, the Sagittarius stream) may be more important than the neglect of the LMC.

\citet{CorreaMagnus2022} implemented a more sophisticated distribution function-based dynamical modelling method, which allows for more flexible density and anisotropy profiles. They also introduced an additional step in modelling to compensate the LMC perturbation. For any choice of the Galactic potential and LMC mass, the orbits of the LMC and the Milky Way are reconstructed in the approximation of analytic non-deforming extended bodies (Equation~\ref{eq:orbit_reconstruction}). Then the orbits of tracer objects are rewound back in time, starting from the present-day measured phase-space coordinates, in the combined potential of the Milky Way and moving LMC (including the acceleration from the reflex motion of the Galactocentric reference frame). Finally, the positions and velocities of tracer objects at the time 2--3~Gyr ago are assumed to be drawn from an equilibrium distribution function, whose parameters (together with the parameters of the Milky Way potential) are optimised to match the observations. They demonstrated that this method is able to produce unbiased potential estimates when doing the extra compensatory step, while neglecting the LMC perturbation biases the potential up by $\sim15\%$, in agreement with \citep{Erkal2020b}. Finally, they applied the method to the kinematic sample of globular clusters and satellite galaxies with 6d phase-space coordinates, and determined a value of the Milky Way mass within 100~kpc that is somewhat higher but compatible with the estimates from the smooth stellar halo \citep{Deason2021} and streams \citep{Vasiliev2021, Koposov2023} (only the studies accounting for the LMC perturbation are listed). A possible reason for a higher estimated mass is that if the population of tracers (satellite galaxies) is incomplete and misses objects near apocentres of their orbits, this biases the derived mass up (see Appendix A in \citep{CorreaMagnus2022}). They also derived constraints on the LMC mass that are comparable to those from stream modelling (Section~\ref{sec:local_effects}, Figure~\ref{fig:lmc_mass}).

Interestingly, the 10--20\% reduction in the estimated Milky Way mass when accounting for the LMC is nearly cancelled when adding back the LMC mass itself (which makes sense given that it is almost entirely within the Galactic virial radius at present). It is natural to expect that the bias would be proportional to the LMC mass, but the fact that it matches it even quantitatively is probably coincidental and might depends on details of the LMC orbit and its phase.

\begin{figure}
\includegraphics[width=14cm]{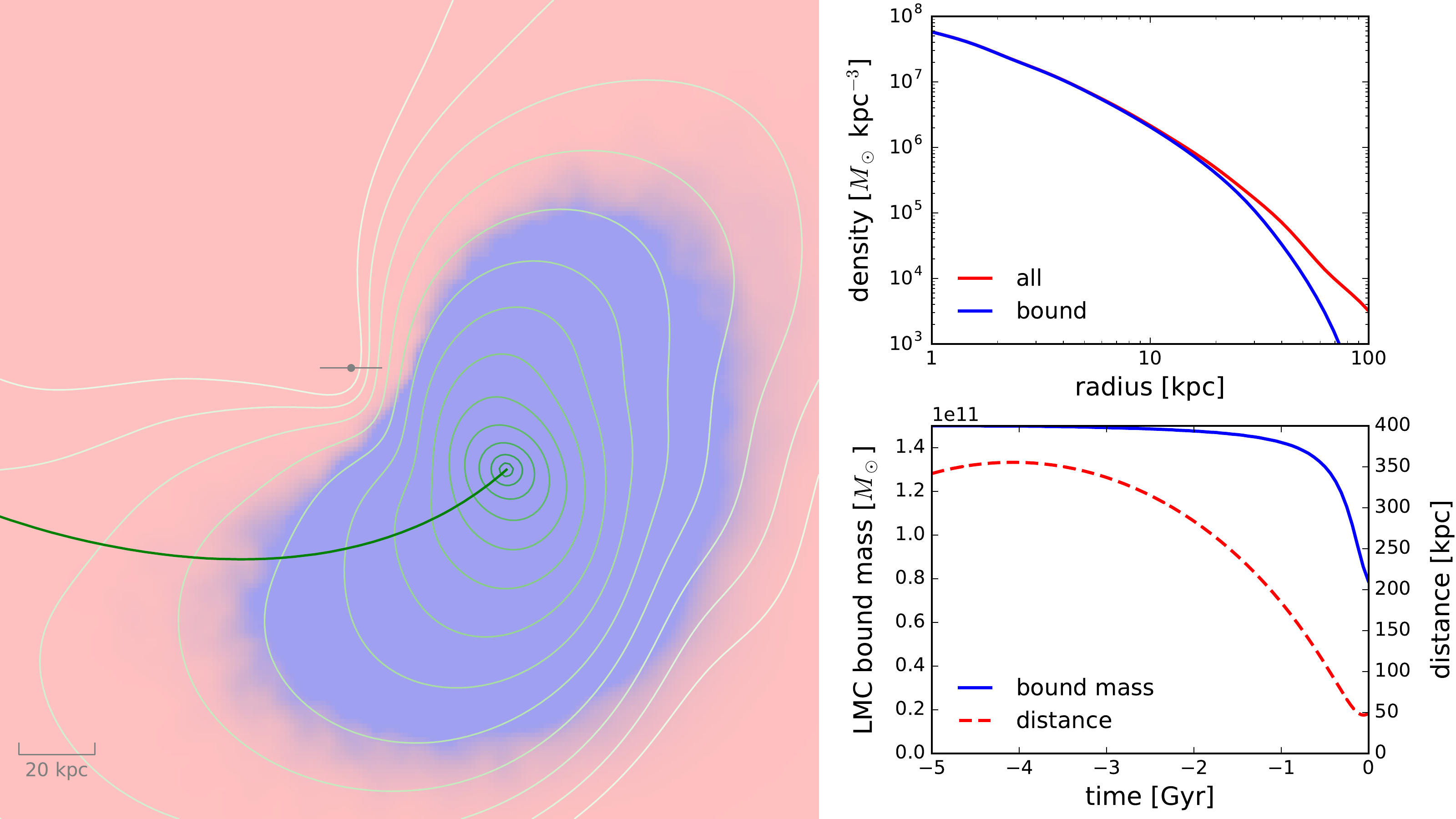}
\caption{Illustration of tidal deformation and mass loss of the LMC. Left panel shows the projected density contours of the LMC particles at present time in the same orientation as in Figure~\ref{fig:model}, top panel. Inner blue-shaded region contains particles still bound to the LMC (having negative energy in the associated reference frame), while particles in the outer red-shaded region are unbound. The Milky Way disc is shown by a gray line with the dot marking its centre, and the past orbit of the LMC is shown by a green curve. Top right panel shows the spherically-averaged density profile of all (red) and bound (blue) particles, with the bound fraction dropping precipitously beyond 30--40 kpc from the LMC centre. Bottom right panel shows the time evolution of bound mass (blue) and Galactocentric distance (red) of the LMC.
}  \label{fig:boundmass}
\end{figure}

\new{Finally, we note that almost all studies quoted the LMC mass before its infall into the Milky Way, but not all of this mass remains bound at present. The concept of bound mass is ill-defined in the case of a highly dynamic interaction, because the gravitational potential changes rapidly and it is generally not possible to determine if a given particle will remain attached to the satellite after it reaches the next apocentre of its orbit. A plausible recipe is to declare a particle to be bound if it has negative total energy in the reference frame of the moving LMC, $E_i=\Phi_{\rm LMC}(\boldsymbol{x}_i) + \frac{1}{2}|\boldsymbol{v}_i-\boldsymbol{v}_{\rm LMC}|^2$, where $\Phi_{\rm LMC}$ is the potential created only by bound LMC particles and does not include the Milky Way. As this definition is self-referencing, the procedure needs to be repeated several times until the list of bound particles stabilises. Figure~\ref{fig:boundmass} demonstrates that the bound mass starts to rapidly decrease once the LMC comes within 100 kpc from the Milky Way $\sim 0.5$~Gyr ago. Particles beyond a few tens kpc from the LMC centre are no longer bound to it and form tidal tails, although they still contribute to the gravitational perturbation of the Milky Way.}

\subsection{Effects on the Local Group}  \label{sec:local_group}

The reflex velocity that the inner Milky Way acquires with respect to the outer halo has consequences for the dynamics of the entire Local Group (or better say, affects our inference about these processes from the present-day velocity measurements of various objects, which has been distorted by our motion induced by the LMC). For instance, the past orbits of Milky Way satellites computed from their current phase-space coordinates, assuming the perfect knowledge of the Galactic potential but ignoring the LMC, may be rather different from their true orbits in the presence of the LMC. On the other hand, if we can reconstruct sufficiently reliably the past orbits of the Milky Way and the LMC for any choice of their potential profiles, then the computation of orbits of other satellites becomes a straightforward exercise -- marred, as usual, by the uncertainties in their current phase-space coordinates.

The most obvious effect of including the LMC is on orbits of its own satellites, which have been discussed in Section~\ref{sec:lmc_satellites}. 
After \textit{Gaia} DR2 made possible the measurement of PM for most dwarf galaxies in the Milky Way neighbourhood, several studies provided the census of likely LMC satellites, based either on the proximity of their angular momentum plane to that of the LMC \citep{Kallivayalil2018,Fritz2019} or on a negative total energy in the LMC-centered reference frame for a significant interval of time \citep{Erkal2020a,Patel2020}. Because of uncertainties in the present-day phase-space coordinates, orbits of even genuine LMC satellites cannot be accurately reconstructed beyond a few Gyr \citep{DSouza2022}, but if a dwarf galaxy has been gravitationally bound to the LMC for at least a Gyr in the majority of orbit samples, it can be considered a LMC satellite with high confidence.

\citet{Battaglia2022} provided an updated catalogue of PM of almost all known Milky Way satellites using \textit{Gaia} eDR3 and then used these data to reconstruct their past orbits for two choices of static Milky Way potentials and a time-dependent Milky Way + LMC potential from \citep{Vasiliev2021}. They confirmed the association with the LMC for a number of dwarfs, but also demonstrated that orbits of many Milky Way satellites are significantly changed in the presence of the LMC (their Appendix~D). This comparison was limited to a single choice of Milky Way potential and just one realization of orbit for each object. \citet{CorreaMagnus2022} performed a much more detailed study of the effect of the LMC on satellite orbits, marginalizing over the uncertainties in their present-day phase-space coordinates and exploring a range of Milky Way potentials and LMC masses. Their Figure~11 illustrates the variety of outcomes: some galaxies are hardly affected by the LMC, others increase or decrease their orbital energy and correspondingly have longer or shorter periods when their orbits are corrected for the LMC influence. For instance, Draco, Sculptor and Ursa Minor have been on less bound orbits until the arrival of the LMC, whereas Leo~I had a lower energy. We stress that in most cases, these changes are driven not by a close encounter with the LMC itself, but by modifying the Milky Way's velocity relative to that of the satellite. Leo~I is a particularly noteworthy example, as its current Galactocentric radial velocity likely exceeds the local escape speed, unless the Milky Way is very massive. In the past, its high recession speed was invoked as an argument in favour of a heavy Galaxy (significantly more massive than preferred by most other techniques), but if it is reduced by a few tens \kms on account of the LMC displacement of our Galaxy, this eases the tension in the Milky Way mass estimates, as already noted by \citet{Erkal2020b}. As discussed in the previous section, accounting for the LMC perturbation also slightly reduces the mass profile of our Galaxy inferred from the ensemble of satellites as kinematic tracers.

It is well known that a large fraction of dwarf galaxies around the Milky Way (including the LMC) share a common plane, dubbed the ``Vast Polar Structure'' (VPOS) \citep{Kroupa2005,Pawlowski2012}, and the PM provided by \textit{Gaia} confirm that orbits of most of these objects also lie in this relatively thin plane \citep{Fritz2018,Pawlowski2020}, even after excluding the satellites of the LMC. \citet{GaravitoCamargo2021b} suggested that the LMC-induced perturbation of satellite orbits may bring them closer to the alignment \new{due to several factors: the direct action of the massive LMC, the kinematic offset of the outer halo with respect to the observer in the inner Milky Way, and the associated deformations in the Galactic gravitational potential (both the local wake and the global dipole perturbation). Other studies \citep{Pawlowski2022,CorreaMagnus2022} found the effect to be rather weak and insufficient to explain the observed degree of clustering of orbital planes, not least because many of these galaxies are fairly distant to be significantly affected. However, a comprehensive analysis of this phenomenon using suitably chosen Milky Way--LMC analogues from cosmological simulations has not yet been performed.}

\begin{figure}
\includegraphics{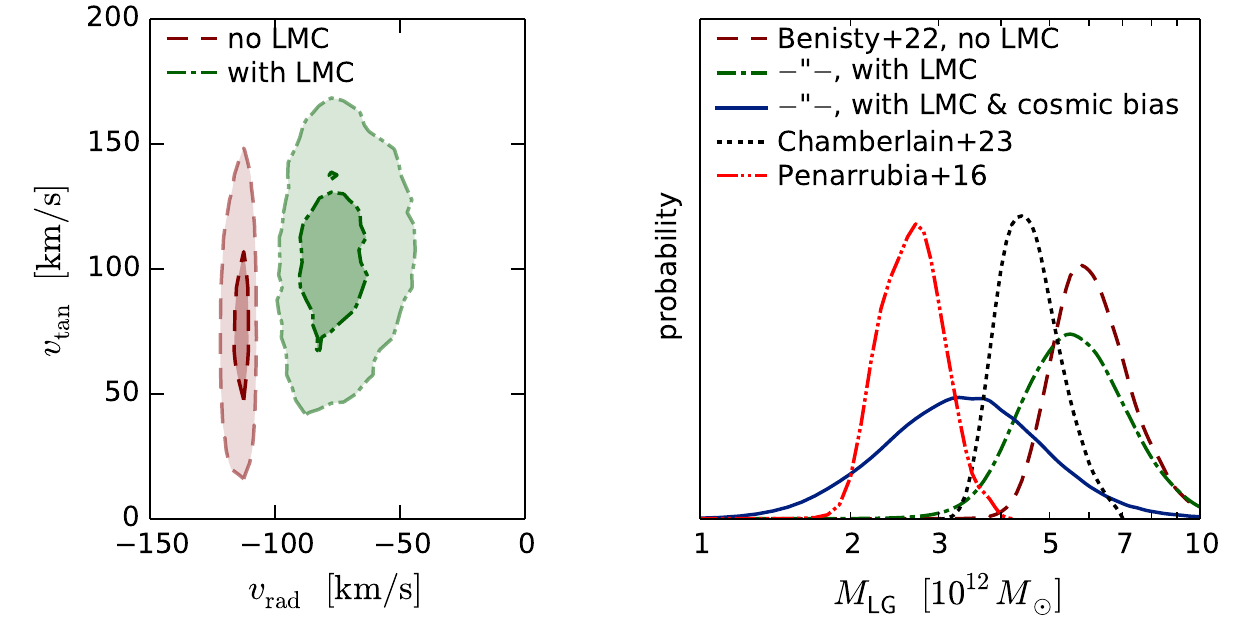}
\caption{\textit{Left panel:} relative velocity of Andromeda and Milky Way. Red contours show the measured values (using PM from \citep{Salomon2021}), green -- corrected for the LMC-induced motion of the Milky Way centre. \textit{Right panel:} inferred mass of the Local Group (or more correctly, Milky Way and Andromeda) from the timing argument \citep{Benisty2022}. Red dashed line is the result using the measured relative velocity, green dot-dashed line -- after correcting for the LMC perturbation, blue solid -- additionally multiplying by a cosmic bias correction factor. Black dotted and red dash-double-dotted lines show results of \citep{Chamberlain2023} and \citep{Penarrubia2016}, who only account for the LMC but not the cosmic bias.
}  \label{fig:timingargument}
\end{figure}

Finally, the reflex motion of central parts of our Galaxy induced by the LMC has implications for the so-called ``timing argument'' -- the estimate of the total mass of the Local Group (dominated by the Milky Way and Andromeda) from their orbital period. The argument \citep{Kahn1959} postulates that since Andromeda is currently approaching the Milky Way, it must have completed more than a half of its orbital period since the Big Bang. Assuming that the (proto-)galaxies were at the pericentre of a Keplerian orbit 13.7~Gyr ago, one can infer the sum of their masses by matching the orbital solution to the present-day distance and relative velocity of Andromeda w.r.t.\ the Milky Way. The radial component of this velocity (almost identical to the line-of-sight velocity of Andromeda corrected for Solar motion around the Galactic centre) is $-114\pm1$~\kms, and the tangential velocity has been measured with increasing precision by \textit{HST} and \textit{Gaia} \citep{vanderMarel2019,Salomon2021} in the range $30$--$100$~\kms. However, both these velocity components have contributions from the LMC-induced reflex motion of the Milky Way. 
\citet{Penarrubia2014} demonstrated that the velocities of galaxies within a few Mpc closely follow the Hubble law with a constant offset proportional to the mass of the Local Group, \new{and determined the latter to be $(2.3\pm0.7)\times10^12\,M_\odot$.} 
\citet{Penarrubia2016} further found that the scatter of measured velocities about this mean trend is minimized when the radial velocity of Andromeda w.r.t.\ the Milky Way is corrected for the reflex velocity induced by a fairly massive LMC ($\sim 2.5\times10^{11}\,M_\odot$), \new{which also decreases the inferred mass of Milky Way + Andromeda and brings it into better agreement with the previous estimate from the local Hubble flow.}
Undoing this perturbation, the radial velocity of Andromeda should be reduced in magnitude, while the tangential velocity increases (Figure~\ref{fig:timingargument}, left panel); to first order, these changes are proportional to the LMC mass. 
\citet{Benisty2022} revisited the classical timing argument, taking into account several factors: (1) correction of apparent radial and tangential velocity of Andromeda for the LMC perturbation (using the approach of \citep{CorreaMagnus2022}), (2) effect of dark energy on the cosmological expansion, and (3) cosmic bias correction factor for the total mass of the Local Group calibrated against cosmological simulations (i.e., applying the timing argument to pairs of galaxies resembling Milky Way and Andromeda, and determining the multiplicative factor that brings the obtained result into agreement with the actual masses of these galaxies). They find that the inclusion of the LMC reduces the inferred Local Group mass by $\lesssim 10\%$, whereas the cosmic bias factor, which accounts for the approximate nature of the timing argument itself, reduces the mass by 40\%.
\citet{Chamberlain2023} performed a similar analysis while only accounting for the LMC perturbation (which also had a $\sim10\%$ effect), and obtained a narrower posterior distribution, but compatible with other estimates except \citep{Penarrubia2016}, who had a similarly narrow distribution but shifted to even lower masses. These estimates are summarised in the right panel of Figure~\ref{fig:timingargument} and are broadly consistent with the sum of dynamically inferred masses of both galaxies.

\section{Conclusions}

The Magellanic Clouds are not only spectacular and inspiring, but also quite unusual -- only a small fraction of Milky Way-like galaxies in the local Universe have a comparably large pair of satellites \citep{Liu2011,Tollerud2011}. Even more peculiar is that they are currently just past the pericentre of their Galactic orbit, which amplifies their dynamical effects on the Milky Way. Although still ranked as a minor merger and remaining a perturbation, the dynamical consequences of this ongoing encounter between our Galaxy and the LMC are significant and well detectable in the currently available observational data. As the precision of our measurements keeps improving, astronomers will be facing even more ambitious challenges in exploring the structure and evolution of the outer regions of Milky Way, for instance, detecting the traces of earlier accretion events or constraining the 3d shape of the Galactic halo. Obviously, the impact of the LMC cannot be ignored at the level of detail warranted by these studies.

The following aspects of the LMC--Milky Way interaction are worth emphasising:
\begin{itemize}
\item The total pre-infall mass of the LMC is likely to be (1--2)${}\times10^{11}\,M_\odot$, i.e., only 5--10 times smaller than the Milky Way mass. This estimate is supported by a number of empirical arguments reviewed in Section~\ref{sec:lmc_mass}, and the dynamical effects of the LMC on the Milky Way discussed in Section~\ref{sec:dynamics} are best explained by a similar mass range. If the LMC is on its first passage around the Galaxy, its dark halo becomes deformed but is still almost entirely brought within 100~kpc from the Milky Way centre.
\item Despite the increasing precision of PM measurements, there remains a considerable uncertainty in the orbital period and apocentre distance of the LMC (Section~\ref{sec:past_orbit}). This is largely due to the orbit being only marginally (if at all) bound to the Milky Way. Likewise, these parameters depend strongly on the Milky Way mass profile. The evidence for the first-passage scenario is less strong now than it was 15 years ago \cite{Besla2007}, since the most recent PM measurements reduce the tangential velocity by a few tens \kms. The orbital period most likely exceeds 5~Gyr, unless the Milky Way is significantly more massive than $10^{12}\,M_\odot$ (which is disfavoured by current models), but an earlier pericentre passage at a distance of $\sim100$~kpc cannot be ruled out with certainty. The implications of this alternative second-passage scenario for the LMC satellites, the Magellanic Stream, and the Milky Way itself are poorly studied.
\item The most obvious dynamical consequences of the massive LMC are local perturbations to objects that pass in its vicinity (Section~\ref{sec:local_effects}), but equally if not more important is its global effect on the Milky Way (Section~\ref{sec:global_effects}). It is often associated with the reflex motion of the Galaxy about the common centre of mass of the Milky Way--LMC system, but this is only part of the story. 
Stars and other objects in the outer halo of the Milky Way (roughly beyond 30~kpc) are not displaced by the LMC in the same way as the inner Galaxy; in other words, the differential perturbation causes a deformation of the Milky Way both in space and in kinematics. This phenomenon only began to be appreciated in the last few years, and is now clearly seen both in simulations and in observations.
\item The LMC-induced reflex motion of the Milky Way leads to an overall reduction of its inferred past orbital period in the case of a more massive LMC compared to a test-particle orbit. This counteracts the more well-known effect of dynamical friction, and has not been accounted for in many earlier studies.
\end{itemize}

\new{Despite significant progress in recent years, several open questions and associated challenges need to be addressed by future studies:
\begin{itemize}
\item As already mentioned, the reconstruction of the past orbit of the LMC is still uncertain and very sensitive to variations in its present-day phase-space coordinates, parameters of the Milky Way potential and other factors. 
\item Connected to the previous point, any modelling effort that aims at exploring the dependence of the past LMC orbit on its mass or on the Milky Way potential must match the present-day position and velocity of the LMC with very high accuracy (better than 1~kpc and a few \kms) to ensure a meaningful comparison between different cases. This is nearly impossible to achieve in large-volume cosmological simulations and is very difficult even in dedicated simulations of the Milky Way--LMC[--SMC] system; this level of precision was rarely attained or even mandated in previous studies.
On the other hand, the LMC trajectory in the last few hundred Myr (up to 100--150~kpc from the Galactic centre) is only weakly sensitive to its current velocity or the Galactic potential, so most of the dynamical effects on the Milky Way do not strongly depend on these factors (but do scale with the LMC mass).
\item Given that full $N$-body simulations of the Milky Way--LMC interaction are expensive and difficult to conduct with sufficient precision, alternative computationally cheaper methods need to be more accurate and sophisticated. The classical dynamical friction expression poorly describes the orbital evolution of massive satellites \citep{Vasiliev2022} even after a manual tuning of the Coulomb logarithm, and the distortions in the gravitational potential of both galaxies have non-negligible dynamical effects \citep{Lilleengen2023}; these factors are ignored in the popular approximation of orbital evolution of two extended but non-deforming bodies.
\item Although the observed spatial and kinematic signatures of the LMC-induced perturbations in the outer halo qualitatively match the models, there appear to be significant differences in the amplitude and direction of these perturbations \citep{Petersen2021,Conroy2021}.
\item The analysis of the perturbations is hampered by the scarcity of available kinematic tracers: existing major spectroscopic surveys with the exception of SEGUE \citep{Yanny2009} contain very few stars beyond 50~kpc. Fortunately, the sample will expand significantly in the next years with the public data releases from DESI \citep{Cooper2023}, WEAVE \citep{Jin2023} and H3 \citep{Conroy2019} surveys.
\end{itemize}
}

The LMC will continue orbiting the Milky Way and will inevitably merge with it in a couple of Gyr. As discussed by \citet{Cautun2019}, this will bringing about further changes in the properties of our Galaxy, e.g., dramatically increasing the mass and mean metallicity of its stellar halo. By the time this merger is complete, however, future astronomers will face an even bigger calamity -- the impeding collision with Andromeda. 
Regardless of whether the LMC is boon or bane for a current-generation dynamicist, we have to endure its presence for the time being, and better learn to cope with its effects on our Galaxy.


\dataavailability{The snapshots, potentials and other files from $N$-body simulations of the Milky Way--LMC interaction for several choices of the LMC mass and the Galactic potential are provided at this zenodo repository: \url{https://doi.org/10.5281/zenodo.7832266}. The main kinematic features induced by the LMC for any other choice of these parameters (Figure~\ref{fig:model}) can be well reproduced without running a full $N$-body simulation, using test-particle integrations in a time-dependent potential of the moving but non-deforming galaxies \citep{CorreaMagnus2022}. A script is provided at \url{https://github.com/GalacticDynamics-Oxford/Agama/blob/master/py/example_lmc_ mw_interaction.py} }

\acknowledgments{I thank Gurtina Besla, Denis Erkal, Nico Garavito-Camargo, Jorge Pe\~narrubia, Mike Petersen, Simon Rozier, the anonymous referees and other colleagues for valuable discussions and suggestions.}


\begin{adjustwidth}{-\extralength}{0cm}

\reftitle{References}

\bibliography{paper}

\PublishersNote{}
\end{adjustwidth}
\end{document}